\documentclass[final,journal,letterpaper,twoside,twocolumn]{IEEEtran}

\usepackage[utf8]{inputenc} 
\usepackage[T1]{fontenc} 
\usepackage[cmex10]{amsmath} 

\usepackage{amssymb} 
\usepackage{mathtools} 
\usepackage{amsfonts} 
\usepackage{accents} 

\usepackage{mleftright}\mleftright 

\usepackage{multirow}

\usepackage[dvips]{graphicx} 
\graphicspath{{./}}
\DeclareGraphicsExtensions{.eps}

\usepackage[dvipsnames]{xcolor} 

\usepackage{bbm} 
\usepackage{xfrac} 
\usepackage{cancel} 
\usepackage{wrapfig} 
\usepackage{tensor} 

\usepackage{theorem}
\usepackage{cite}

\usepackage{url} 

\usepackage{psfrag} 

\usepackage[inline]{enumitem} 

\usepackage{array} 

\usepackage{aliascnt} 

\usepackage[hidelinks]{hyperref} 

\usepackage{cleveref} 

\usepackage{orcidlink} 

\usepackage[ruled,algoruled,vlined]{algorithm2e}

\SetCommentSty{mycommfont}

\usepackage{tikz}

\usetikzlibrary{calc,intersections,arrows.meta}
\usepackage{pgfplots}
\usepgfplotslibrary{fillbetween}

\usepackage{xparse} 

\usepackage{ifthen} 

\usepackage{comment}

\crefname{section}{Section}{Sections}
\crefname{subsection}{Section}{Sections}
\crefname{equation}{Eq.}{Equations}
\crefname{enumi}{part}{parts}
\crefname{table}{Table}{Tables}
\crefname{figure}{Figure}{Figures}
\crefname{algocf}{Algorithm}{Algorithms}

\newtheorem{theorem}{Theorem}
\crefname{theorem}{Theorem}{Theorems}

\newaliascnt{lemma}{theorem}
\newtheorem{lemma}[lemma]{Lemma}
\aliascntresetthe{lemma}
\crefname{lemma}{Lemma}{Lemmas}

\newaliascnt{definition}{theorem}
\newtheorem{definition}[definition]{Definition}
\aliascntresetthe{definition}
\crefname{definition}{Definition}{Definitions}

\newaliascnt{corollary}{theorem}
\newtheorem{corollary}[corollary]{Corollary}
\aliascntresetthe{corollary}
\crefname{corollary}{Corollary}{Corollarys}

\newaliascnt{claim}{theorem}

\aliascntresetthe{claim}
\crefname{claim}{Claim}{Claims}

\newaliascnt{conjecture}{theorem}

\aliascntresetthe{conjecture}
\crefname{conjecture}{Conjecture}{Conjectures}

\newaliascnt{question}{theorem}

\aliascntresetthe{question}
\crefname{question}{Question}{Questions}

\newaliascnt{oquestion}{theorem}

\aliascntresetthe{oquestion}
\crefname{oquestion}{Open Question}{Open Questions}

\theoremstyle{plain}
\theorembodyfont{\normalfont}

\newtheorem{cnstr}{Construction}

\newenvironment{construction}{\begin{cnstr}}{\hfill$\Box$\end{cnstr}}
\crefname{cnstr}{Construction}{Constructions}

\crefname{step}{Step}{Steps}

\crefname{regime}{Regime}{Regimes}

\newtheorem{myalgo}{Algorithm}

\crefname{myalgo}{Algorithm}{Algorithms}

\newcommand\numberthis{\stepcounter{equation}\tag{\theequation}}

\newcounter{enumrom}
\renewcommand{\theenumrom}{(\roman{enumrom})}

\makeatletter
\renewcommand{\@endtheorem}{\endtrivlist}
\makeatother

\makeatletter
\renewcommand{\thefigure}{{\@arabic\c@figure}}
\renewcommand{\fnum@figure}{{\bf Figure\,\thefigure}}
\makeatother

\renewcommand{\leq}{\leqslant}
\renewcommand{\geq}{\geqslant}

\newcommand{\bfb}{{\boldsymbol b}}
\newcommand{\bfc}{{\boldsymbol c}}

\newcommand{\bff}{{\boldsymbol f}}

\newcommand{\bfh}{{\boldsymbol h}}

\newcommand{\bfu}{{\boldsymbol u}}
\newcommand{\bfv}{{\boldsymbol v}}
\newcommand{\bfw}{{\boldsymbol w}}
\newcommand{\bfx}{{\boldsymbol x}}
\newcommand{\bfy}{{\boldsymbol y}}
\newcommand{\bfz}{{\boldsymbol z}}

\newcommand{\cC}{\mathcal{C}}

\newcommand{\cF}{\mathcal{F}}

\newcommand{\cL}{\mathcal{L}}

\newcommand{\cR}{\mathcal{R}}

\newcommand{\cT}{\mathcal{T}}

\newcommand{\cX}{\mathcal{X}}

\renewcommand{\Bbb}{\mathbb}

\newcommand{\N}{{\Bbb N}}

\DeclarePairedDelimiter\abs{\lvert}{\rvert}
\DeclarePairedDelimiter\ceilenv{\lceil}{\rceil}
\DeclarePairedDelimiter\floorenv{\lfloor}{\rfloor}
\DeclarePairedDelimiter\parenv{\lparen}{\rparen}

\DeclarePairedDelimiter\bracenv{\lbrace}{\rbrace}
\DeclarePairedDelimiterX\mathset[2]{\lbrace}{\rbrace}{#1 : #2}
\DeclarePairedDelimiterX\inner[2]{\langle}{\rangle}{#1 \mathrel{},\mathrel{} #2}
\DeclarePairedDelimiterX\condparenv[2]{(}{)}{#1 \mathrel{}\delimsize\vert\mathrel{} #2}

\newcommand{\mybinom}[2]{\Biggl(\begin{array}{@{}c@{}}#1\\#2\end{array}\Biggr)}

\DeclareDocumentCommand\norm{ o m }{
    \IfNoValueTF{#1}
        {\left\Vert#2\right\Vert}
        {\left\Vert#2\right\Vert_{#1}}
}

\DeclareDocumentCommand\der{ o m o }{
    \IfNoValueTF{#1}
        {
            \IfNoValueTF{#3}
                {\frac{d}{d{#2}}}
                {\frac{d{#3}}{d{#2}}}
        }
        {\parenv*{\frac{d}{d{#2}}}^{#1}\IfNoValueTF{#3}{}{#3}}
}
\DeclareDocumentCommand\partder{ o m m }{
    \IfNoValueTF{#1}
        {\frac{\partial{#3}}{\partial{#2}}}
        {\frac{\partial^{#1}{#3}}{{\partial{#2}}^{#1}}}
}
\DeclareDocumentCommand\df{ o m o }{
    d\IfNoValueTF{#1}{}{^{#1}}{#2}\IfNoValueTF{#3}{}{_{#3}}
}

\newcommand{\deq}{\mathrel{\triangleq}}

\DeclareMathOperator{\rll}{\cR\cL\cL}

\DeclareMathOperator{\red}{red}

\DeclareMathOperator{\rf}{\cR\cF}

\newcommand{\lmin}{\ell_{\min}}

\newcommand{\lover}{\ell_{\operatorname*{over}}}

\DeclareDocumentCommand\enc{ o }{
    \IfNoValueTF{#1}
        {\operatorname{Enc}}
        {\operatorname{Enc}_{\ref*{#1}}}
}

\DeclareDocumentCommand\dec{ o }{
    \IfNoValueTF{#1}
        {\operatorname{Dec}}
        {\operatorname{Dec}_{\ref*{#1}}}
}

\makeatletter
\newcommand\code[1]{%
  \@ifundefined{r@#1}{%
    \cC_{\operatorname*{#1}}%
  }{%
    \cC_{\ref*{#1}}%
  }%
}
\makeatother

\begin{document}

\title{Generalized Unique Reconstruction from Substrings}

\author{Yonatan~Yehezkeally\,\orcidlink{0000-0003-1652-9761}\,%
		,~\IEEEmembership{Member,~IEEE}, 
		Daniella~Bar-Lev\,\orcidlink{0000-0001-6766-1450}\,%
		,~\IEEEmembership{Student~Member,~IEEE}, 
		Sagi~Marcovich\,\orcidlink{0000-0003-4165-2024}\,%
		,~\IEEEmembership{Student~Member,~IEEE} 
		and Eitan~Yaakobi\,\orcidlink{0000-0002-9851-5234}\,%
		,~\IEEEmembership{Senior~Member,~IEEE}
  \thanks{Manuscript received October 10, 2022; revised February 23, 2023; accepted April 16, 2023. 
    This work was supported in part by the European Research Council (ERC) through the European Union's Horizon 2020 Research and Innovation Programme under Grant 801434 and by the Israel Innovation Authority Grant 75855. 
  Y.~Yehezkeally was supported by a Carl Friedrich von Siemens postdoctoral research fellowship of the Alexander von Humboldt Foundation. 
  D.~Bar-Lev, S.~Marcovich, and E.~Yaakobi were supported in part by the United States-Israel BSF grant no. 2018048. 
  This article was presented in part at the 2021 {IEEE} Information Theory Workshop [\textsc{DOI:\; 10.1109/ITW48936.2021.9611486}], and at the 2022 International Symposium on Information Theory and Its Applications. 
  \emph{(Corresponding author: Yonatan Yehezkeally)}}%
  \thanks{%
  Y.~Yehezkeally is with the Institute for Communications Engineering, School of Computation, Information and Technology, Technical University of Munich, 80333 Munich, Germany 
  (e-mail: \texttt{yonatan.yehezkeally@tum.de}).
  D.~Bar-Lev, S.~Marcovich, and E.~Yaakobi are with the Department of Computer Science, Technion---Israel Institute of Technology, Haifa 3200003, Israel 
  (e-mails: \texttt{\{yaakobi,daniellalev,sagimar\}@cs.technion.ac.il}).}%
  \thanks{%
  The first three authors contributed equally to this work.}%
  \thanks{Copyright (c) 2023 IEEE. Personal use of this material is permitted.  Permission from IEEE must be obtained for all other uses, in any current or future media, including reprinting/republishing this material for advertising or promotional purposes, creating new collective works, for resale or redistribution to servers or lists, or reuse of any copyrighted component of this work in other works.}
}

\maketitle
\begin{abstract}
This paper introduces a new family of reconstruction codes which is 
motivated by applications in DNA data storage and sequencing. In such 
applications, DNA strands are sequenced by reading some subset of 
their substrings. While previous works considered two extreme cases in 
which \emph{all} substrings of pre-defined lengths are read or 
substrings are read with no overlap for the single string case, this 
work studies two extensions of this paradigm. The first extension 
considers the setup in which consecutive substrings are read with some 
given minimum overlap. First, an upper bound is provided on the 
attainable rates of codes that guarantee unique reconstruction. Then, 
efficient constructions of codes that asymptotically meet that upper 
bound are presented. In the second extension, we study the setup where 
multiple strings are reconstructed together. Given the number of 
strings and their length, we first derive a lower bound on the read 
substrings' length~$\ell$ that is necessary for the existence of 
multi-strand reconstruction codes with non-vanishing rates. We then 
present two constructions of such codes and show that their rates 
approach~1 for values of~$\ell$ that asymptotically behave like the 
lower bound.
\end{abstract}

\section{Introduction} \label{sec:intro}

String reconstruction refers to a large class of problems where 
information about a string can only be obtained in the form of 
multiple, incomplete and/or noisy observations. 
Examples of such problems are the \emph{reconstruction problem} by 
Levenshtein~\cite{Lev01b}, the \emph{trace reconstruction 
problem}~\cite{BatKanKhaMcG04,CheGabMilRib20}, and the \emph{$k$-deck 
problem}~\cite{BenMeySchSmiSto91, Sco97, DudSch03, 
ChrKiaRaoVarYaaYao19}. 

Notably, when observations are comprised of unordered consecutive 
substrings, two distinct models have received significant interest in 
the past decade due to applications in DNA- or polymer-based storage 
systems, resulting from contemporary sequencing 
technologies~\cite{MotBreTse13, BreBreTse13, GabMil19}. 
The first is the \emph{reconstruction from substring-compositions 
problem}~\cite{MotBreTse13, MotRamTseMa13, BreBreTse13, 
AchDasMilOrlPan15, ShoCouTse15, GanMosRac16, ShoKamGovXiaCouTse16, 
KiaPulMil16, EliGabMedYaa21, MarYaa21, YehPol21} (including extensions 
for erroneous observations~\cite{MarYaa21, GabMil19, ChaChrEzeKia17, 
YehPol21}), which arises from an idealized assumption of full overlap 
(and uniform coverage) in read substrings; the second is the 
\emph{torn-paper problem}~\cite{RavVahSho21, ShoVah21, NasShoVah22, 
BarMarYaaYeh22} (a problem closely related to the \emph{shuffling 
channel}~\cite{ShoHec19, HecShoRamTse17, LenSieWacYaa19, WeiMer22}), 
which results from an assumption of no overlap. In applications, the 
distinction models the question of whether the complete information 
string may be replicated and uniformly segmented for sequencing, or if 
segmentation occurs adversarially in the medium prior to sequencing.

Motivated by these two paradigms, we study in this paper a generalized 
(or intermediate) setting where an information string is observed 
through an arbitrary collection of its substrings, where the minimum 
length of each retrieved substring, as well as the length of overlap 
between consecutive substrings, are bounded from below. 
A similar setting was recently studied in~\cite{RavVahSho22}, where 
both substrings' lengths and overlap were assumed to be random; we 
study the problem in the aforementioned worst-case regime.

Further, in both sequencing and tandem-mass-spectrometry 
technologies, used for DNA and polymer-based storage systems 
respectively, it is typical that not a single string is read alone, 
but multiple strings simultaneously~\cite{Sal10, Chinetal13, 
LomQuiSim15, KhaPerBabNavSho18, GabPatMil23}. We therefore study a 
setting where retreived substrings are taken from a collection of 
information strings stored together, with no information on the string 
from which they originated. We remark that this extension was already 
studied by the authors for the torn-paper problem, 
in~\cite{BarMarYaaYeh22}.

Our problem setting is therefore given as follows: a multiset of~$k$ 
length\nobreakdash-$n$ strings is transmitted, and substrings of all 
information strings are retrieved, such that the length of each 
substring is at least~$\lmin$, and consecutive substrings of the same 
information string overlap in at least~$\lover$ positions. 
We are interested in the minimum value of~$\lmin$, as a function 
of~$k$ and~$n$, for which there exist codes allowing for unique 
reconstruction in this channel with asymptotically non-vanishing 
rates, and then what is the asymptotically optimal obtainable rate 
given the value of~$\lover$. In these cases, we seek to develop 
efficient coding schemes which asymptotically attain optimal rates.

The rest of this paper is organized as follows. In~\cref{sec:def}, we 
present notation and definitions which are used throughout the paper. 
In~\cref{sec:rf}, we overview and extend results in existing 
literature which already solve our problem setting in specific 
end-cases. In~\cref{sec:partial-single,sec:partial-single-encode}, 
we present a solution to the aforementioned problem in the private 
case of a single string~($k=1$), by respectively bounding from above 
the asymptotically attainable rate of codes for unique reconstruction 
as a function of~$\lmin,\lover$, and then developing efficient 
encoding and decoding algorithms for such codes, asymptotically 
meeting this bound. Then, in~\cref{sec:multi}, we study solutions to 
the problem in a different private case, where $\lover=\lmin-1$ (i.e., 
a multi-strand extension of the reconsturction from substrings 
problem); we likewise present bounds and two efficient constructions 
of multiset-codes for this case, whose rates asymptotically 
approach~$1$ for values of~$\lmin$ asymptotically equivalent to the 
lower bound. We conclude in~\cref{sec:conc} with a summary and closing 
remarks.

\section{Definitions and Preliminaries}\label{sec:def}

Let $\Sigma$ be a finite alphabet of size~$q$. Where advantageous, we 
assume $\Sigma$ is equipped with a ring structure, and in particular 
identify elements $0,1\in\Sigma$. For a positive integer~$n$, let 
$[n]$ denote the set $[n] \deq \{0,1,\ldots, n-1\}$. We denote a 
\emph{multiset} by $S = \bracenv*{\bracenv*{a,a,b,\ldots}}$; i.e., 
elements are allowed to appear with multiplicity. For convenience we 
let $\norm{S}$, for a multiset~$S$, denote the number of unique 
elements in $S$.

For two non-negative functions~$f,g$ of a common variable~$n$, 
denoting $L\deq \limsup_{n\to\infty}\frac{f(n)}{g(n)}$ (in the wide 
sense) we say that $f=o_n(g)$ if $L=0$, $f=\Omega_n(g)$ if $L>0$, 
$f=O_n(g)$ if $L<\infty$, and $f=\omega_n(g)$ if $L=\infty$. 
We say that $f=\Theta_n(g)$ if $f=\Omega_n(g)$ and $f=O_n(g)$. 
If $f$ is not positive, we say $f=O_n(g)$ ($f=o_n(g)$) if $\abs*{f} 
= O_n(g)$ (respectively, $\abs*{f}=o_n(g)$). 
If clear from context, we omit the subscript from aforementioned 
notations.

Let $\Sigma^*$ denote the set of all finite strings over $\Sigma$. 
The length of a string~$\bfx = (x_0,x_1,\ldots,x_{n-1})\in \Sigma^*$ 
is denoted by $\abs*{\bfx} = n$. 
For strings $\bfx,\bfy\in\Sigma^*$, we denote their concatenation by 
$\bfx\circ\bfy$. We say that $\bfv$ is a \emph{substring} of $\bfx$ if 
there exist strings $\bfu,\bfw$ such that $\bfx = \bfu\circ \bfv\circ 
\bfw$. If $\bfu$ (respectively, $\bfw$) is empty, we say that $\bfv$ 
is a \emph{prefix} (\emph{suffix}) of $\bfx$. If the length of $\bfv$ 
is $\ell$, we specifically say that $\bfv$ is an 
\emph{$\ell$-substring} of $\bfx$ (similarly, an 
$\ell$-prefix/suffix). 
For $I\subseteq [\abs*{\bfx}]$, we let $\bfx_I$ denote the 
\emph{subsequence} of~$\bfx$ obtained by restriction to the 
coordinates of~$I$ (i.e., when $\bfx$ is considered as a function from 
$[\abs*{\bfx}]$ into $\Sigma$); specifically, for $i\in [\abs*{\bfx}-
\ell+1]$ we denote by $\bfx_{i+[\ell]}$ the $\ell$-substring of~$\bfx$ 
at \emph{location}~$i$ (we reserve the term \emph{index} for a 
different use), where $i+[\ell] = \mathset*{i+j}{j\in [\ell]}$.

We define 
\begin{align*}
	\cX_{n,k} &\deq 
	\mathset*{S=\bracenv*{\bracenv*{\bfx_1,\ldots,\bfx_k}}}{\forall i, 
	\bfx_i\in \Sigma^n},
\end{align*}
and observe that $\abs*{\cX_{n,k}} = \binom{k+q^n-1}{k}$. 
We consider in this paper the problem of multi-string reconstruction 
from substrings with partial overlap. That is, we assume that a 
message $S\in \cX_{n,k}$ is observed only through a multiset of 
substrings of its elements, without order or information on the 
substring from which they originate, with the following restrictions: 
\begin{enumerate*}[label=(\roman*)]
\item all observed substrings are of length at least~$\lmin$; and
\item succeeding substrings of the same~$\bfx\in S$ overlap with 
length at least~$\lover$ (in particular, every symbol of $\bfx$ is 
observed in some substring).
\end{enumerate*}

More formally, a \emph{substring-trace} of $\bfx\in\Sigma^n$ is a 
multiset $\bracenv*{\mathset*{\bfx_{i_j+[\ell_j]}}{1\leq j\leq m}}$, 
for some $m\in\N$, such that $i_1<i_2<\cdots<i_m$ and $\ell_j\in 
[n-i_j+1]$. A substring-trace is \emph{complete} if $i_1=0$, $i_{j+1} 
< i_j+\ell_j$ for all $j<m$, and $i_m+\ell_m=n$. A complete 
substring-trace of $\bfx\in\Sigma^n$ is called an \emph{$(\lmin, 
\lover)$-trace} if $\ell_j\geq \lmin\geq \lover$ for all $j$, and 
$i_j + \ell_j - i_{j+1} \geq \lover$ for all $j < m$. For example, for 
$\bfx = 11101110101111$ 
\begin{itemize}
\item $\bracenv*{\bracenv*{1110111,111010,101111}}$ is a $(6,2)$-trace 
of~$\bfx$; 

\item $\bracenv*{\bracenv*{111011,110101,101111}}$ is a complete 
substring-trace of~$\bfx$ which is not a $(6,2)$-trace; and 

\item $\bracenv*{\bracenv*{110111,110101,01111}}$ is a substring-trace 
of~$\bfx$ which is not complete (since $i_1>0$). 
\end{itemize}
See \cref{fig:exm-1,fig:exm-2,fig:exm-3} for an illustration if these 
substring-traces.
\begin{figure}[t]{}%
\centering
\begin{tikzpicture}[long dash/.style={dash pattern=on 10pt off 2pt},short dash/.style={dash pattern=on 5pt off 1pt}]
\foreach \i in {0,1,2,4,5,6,8,10,11,...,13}{
  \node (x\i) at ($\i*(10pt,0)$){$1$};
}
  \foreach \i in {3,7,9}{
  \node (x\i) at ($\i*(10pt,0)$){$0$};
}
\draw[semithick,short dash,name path=orange,fill=orange,fill opacity=0.3] plot[smooth cycle] 
coordinates {($(x0.north) + (-1pt,2pt)$) ($(x3.north) + (0,2pt)$) ($(x6.north) + (1pt,2pt)$) ($(x6.east) + (0,1pt)$) ($(x6.south) + (1pt,0)$) ($(x3.south) + (0,0)$) ($(x0.south) + (-1pt,0)$) ($(x0.west) + (0,1pt)$)};
\draw[semithick,short dash,name path=green,fill=green,fill opacity=0.3] plot[smooth cycle] 
coordinates {($(x4.north) + (-1pt,0)$) ($(x6.north east) + (0,0)$) ($(x9.north) + (1pt,0)$) ($(x9.east) + (0,-1pt)$) ($(x9.south) + (1pt,-2pt)$) ($(x7.south west) + (0,-2pt)$) ($(x4.south) + (-1pt,-2pt)$) ($(x4.west) + (0,-1pt)$)};
\draw[semithick,short dash,name path=blue,fill=blue,fill opacity=0.3] plot[smooth cycle] 
coordinates {($(x8.north) + (-1pt,2pt)$) ($(x10.north east) + (0,2pt)$) ($(x13.north) + (1pt,2pt)$) ($(x13.east) + (0,1pt)$) ($(x13.south) + (1pt,0)$) ($(x11.south west) + (0,0)$) ($(x8.south) + (-1pt,0)$) ($(x8.west) + (0,1pt)$)};
\end{tikzpicture}
\vspace{-1\baselineskip}
\caption{A $(6,2)$-trace of~$\bfx$.   
\label{fig:exm-1}}
\vspace{1\baselineskip}
\begin{tikzpicture}[long dash/.style={dash pattern=on 10pt off 2pt},short dash/.style={dash pattern=on 5pt off 1pt}]
\foreach \i in {0,1,2,4,5,6,8,10,11,...,13}{
  \node (x\i) at ($\i*(10pt,0)$){$1$};
}
  \foreach \i in {3,7,9}{
  \node (x\i) at ($\i*(10pt,0)$){$0$};
}
\draw[semithick,short dash,name path=orange,fill=orange,fill opacity=0.3] plot[smooth cycle] 
coordinates {($(x0.north) + (-1pt,2pt)$) ($(x2.north east) + (0,2pt)$) ($(x5.north) + (1pt,2pt)$) ($(x5.east) + (0,1pt)$) ($(x5.south) + (1pt,0)$) ($(x3.south west) + (0,0)$) ($(x0.south) + (-1pt,0)$) ($(x0.west) + (0,1pt)$)};
\draw[semithick,short dash,name path=green,fill=green,fill opacity=0.3] plot[smooth cycle] 
coordinates {($(x5.north) + (-1pt,0)$) ($(x7.north east) + (0,0)$) ($(x10.north) + (1pt,0)$) ($(x10.east) + (0,-1pt)$) ($(x10.south) + (1pt,-2pt)$) ($(x8.south west) + (0,-2pt)$) ($(x5.south) + (-1pt,-2pt)$) ($(x5.west) + (0,-1pt)$)};
\draw[semithick,short dash,name path=blue,fill=blue,fill opacity=0.3] plot[smooth cycle] 
coordinates {($(x8.north) + (-1pt,2pt)$) ($(x10.north east) + (0,2pt)$) ($(x13.north) + (1pt,2pt)$) ($(x13.east) + (0,1pt)$) ($(x13.south) + (1pt,0)$) ($(x11.south west) + (0,0)$) ($(x8.south) + (-1pt,0)$) ($(x8.west) + (0,1pt)$)};
\draw[-latex] ($(x5.south) - (5pt,20pt)$) to[out=30,in=270,looseness=1]  ($(x5.south) - (0,5pt)$);
\end{tikzpicture}
\vspace{-1\baselineskip}
\caption{A complete substring-trace of~$\bfx$ (not a $(6,2)$-trace).   
\label{fig:exm-2}}
\vspace{1\baselineskip}
\begin{tikzpicture}[long dash/.style={dash pattern=on 10pt off 2pt},short dash/.style={dash pattern=on 5pt off 1pt}]
\foreach \i in {0,1,2,4,5,6,8,10,11,...,13}{
  \node (x\i) at ($\i*(10pt,0)$){$1$};
}
  \foreach \i in {3,7,9}{
  \node (x\i) at ($\i*(10pt,0)$){$0$};
}
\draw[semithick,short dash,name path=orange,fill=orange,fill opacity=0.3] plot[smooth cycle] 
coordinates {($(x1.north) + (-1pt,2pt)$) ($(x3.north east) + (0,2pt)$) ($(x6.north) + (1pt,2pt)$) ($(x6.east) + (0,1pt)$) ($(x6.south) + (1pt,0)$) ($(x4.south west) + (0,0)$) ($(x1.south) + (-1pt,0)$) ($(x1.west) + (0,1pt)$)};
\draw[semithick,short dash,name path=green,fill=green,fill opacity=0.3] plot[smooth cycle] 
coordinates {($(x5.north) + (-1pt,0)$) ($(x7.north east) + (0,0)$) ($(x10.north) + (1pt,0)$) ($(x10.east) + (0,-1pt)$) ($(x10.south) + (1pt,-2pt)$) ($(x8.south west) + (0,-2pt)$) ($(x5.south) + (-1pt,-2pt)$) ($(x5.west) + (0,-1pt)$)};
\draw[semithick,short dash,name path=blue,fill=blue,fill opacity=0.3] plot[smooth cycle] 
coordinates {($(x9.north) + (-1pt,2pt)$) ($(x11.north) + (0,2pt)$) ($(x13.north) + (1pt,2pt)$) ($(x13.east) + (0,1pt)$) ($(x13.south) + (1pt,0)$) ($(x11.south) + (0,0)$) ($(x9.south) + (-1pt,0)$) ($(x9.west) + (0,1pt)$)};
\draw[-latex] ($(x0.south) - (5pt,15pt)$) to[out=30,in=270,looseness=1]  (x0.south);
\end{tikzpicture}
\vspace{-1\baselineskip}
\caption{An incomplete substring-trace of~$\bfx$.   
\label{fig:exm-3}}
\end{figure}

The \emph{$(\lmin, \lover)$-trace spectrum} of $\bfx\in \Sigma^n$, 
denoted $\cT_{\lmin}^{\lover}(\bfx)$, is the set of all $(\lmin, 
\lover)$-traces of $\bfx$. 
We extend the definition to $S\in \cX_{n,k}$ by 
$\cT_{\lmin}^{\lover}(S)\deq \bigcup_{\bfx\in S} 
\cT_{\lmin}^{\lover}(\bfx)$, where the union respects multiplicity 
(i.e., multiset union), and similarly extend the definitions of 
traces. 
Our channel accepts $S\in \cX_{n,k}$ and outputs a single arbitrary 
$(\lmin, \lover)$-trace of $S$.

For all $\cC\subseteq \cX_{n,k}$, we denote the \emph{rate}, 
\emph{redundancy} of~$\cC$ by $R(\cC)\deq 
\frac{\log\abs*{\cC}}{\log\abs*{\cX_{n,k}}}$, $\red(\cC)\deq 
\log\abs*{\cX_{n,k}}-\log\abs*{\cC}$, respectively. Throughout the 
paper, we use the base-$q$ logarithms. 
Motivated by the above channel definition, a code $\cC\subseteq 
\Sigma^n$ is called an \emph{$(\lmin, \lover)$-trace code} if for 
all $\bfc_1\neq \bfc_2\in \cC$, $\cT_{\lmin}^{\lover}(\bfc_1)\cap 
\cT_{\lmin}^{\lover}(\bfc_2) = \emptyset$. We likewise define a 
\emph{multi-strand $(\lmin, \lover)$-trace code}~$\cC\subseteq 
\cX_{n,k}$. 
The main goal of this work is to find, for $\lmin,\lover$ as functions 
of~$n,k$, the maximum asymptotic rate of (multi-strand) $(\lmin, 
\lover)$-trace codes. We will also be interested in efficient 
constructions of codes with rate asymptotically approaching that 
value.

For convenience of analysis we denote by $\cL_{\lmin}^{\lover}(\bfx) 
\in \cT_{\lmin}^{\lover}(\bfx)$, for $\bfx\in \Sigma^n$, the 
$(\lmin,\lover)$-trace of $\bfx$ containing specifically its 
$\lmin$-prefix, and subsequent $\lmin$-substrings overlapping in 
precisely $\lover$ coordinates. For example, if $\bfx = 
11101110101111$ then 
\begin{align*}
    \cL_4^2(\bfx) = 
    \bracenv*{\bracenv*{1110,1011,1110,1010,1011,1111}}.
\end{align*}
(Here, if $\lmin-\lover$ does not divide $n-\lmin$ we allow the 
$\lmin$-suffix to contain a longer overlap with its preceding 
$\lmin$-substring.) We likewise let $\cL_{\lmin}^{\lover}(S)\deq 
\bigcup_{\bfx\in S} \cL_{\lmin}^{\lover}(\bfx)$.

\section{Repeat-free strings}\label{sec:rf}

In this this section, we discuss the special case of $(\ell, 
\ell-1)$-trace codes, which has been studied in literature in the 
context of reconstruction from substring compositions. To that end, 
we introduce the pertinent notion of \emph{repeat-free} 
strings~\cite{EliGabMedYaa21}, which we denote herein for all $\ell 
\leq n$ by 
\begin{align*}
    \rf_\ell(n)\deq 
    \mathset*{\bfx\in \Sigma^n}{\norm{\cL_\ell^{\ell-1}(\bfx)} = 
    n-\ell+1}, 
\end{align*}
That is, the set of all length-$n$ strings whose $\ell$-substrings are 
all distinct. 
It was observed in~\cite{Ukk1992} that if $\bfx\in \rf_\ell(n)$, 
then $\cL_{\ell+1}^{\ell}(\bfx) \neq \cL_{\ell+1}^{\ell}(\bfy)$ for 
all $\bfy\in \Sigma^n$, $\bfy\neq\bfx$. A straightforward 
generalization of the arguments therein demonstrates the following 
lemma.
\begin{lemma}\label{lem:reconst-rf}
Given $\lmin>\lover$, for all $\bfx\in \rf_{\lover}(n)$, there exists 
an efficient algorithm reconstructing $\bfx$ from any 
$(\lmin,\lover)$-trace of $\bfx$.
\end{lemma}
\begin{IEEEproof}
Let $T$ be any $(\lmin,\lover)$-trace of $\bfx$. For any $\bfu\in T$, 
suppose by negation that there exist $\bfv_1, \bfv_2\in T$, $\bfv_1 
\neq \bfv_2$, such that the $\ell_i$-suffix of $\bfv_i$ equals the 
$\ell_i$-prefix of $\bfu$, where $\ell_i\geq \lover$, for $i\in 
\bracenv*{1,2}$. Since $\bfv_1 \neq \bfv_2$, they occur in distinct 
locations in $\bfx$, and in particular their $\min\bracenv*{\ell_1,
\ell_2}$-suffix occurs in distinct locations; this in contradiction to 
$\bfx\in \rf_{\lover}(n)$. The same argument proves that there do not 
exist $\bfv_1, \bfv_2\in T$, $\bfv_1 \neq \bfv_2$, such that the 
$\ell_i$-prefix of $\bfv_i$ equals the $\ell_i$-suffix of $\bfu$, 
where again $\ell_i\geq \lover$, for $i\in\bracenv*{1,2}$.

Hence, matching prefix to suffix, of lengths at least $\lover$, one 
reconstructs $\bfx$ from $T$. Equivalently, for each $\bfu\in T$, 
finding the unique $\bfv\in T$ that contains the $\lover$-prefix of 
$\bfu$ as a substring (which exists unless $\bfu$ is itself a prefix 
of $\bfx$) results with complete reconstruction. A naive 
implementation requires $O(n^2 \lover)$ run-time. 
\end{IEEEproof}

We also denote \emph{multi-strand $\ell$-repeat-free strings} 
\begin{align*}
    \rf_\ell(n, k)\deq 
    \mathset*{S\in \cX_{n,k}}{\norm{\cL_\ell^{\ell-1}(S)} = 
    k (n-\ell+1)}, 
\end{align*}
and observe the following corollary of \cref{lem:reconst-rf}.
\begin{corollary}\label{cor:reconst-rf-k}
For all $S\in \rf_{\lover}(n,k)$ there exists an efficient algorithm 
reconstructing $S$ from any $(\lmin,\lover)$-trace of~$S$.
\end{corollary}
\begin{IEEEproof}
Observe that $S$ is a set, $S\subseteq \rf_{\lover}(n)$, and that 
$\mathset*{\cL_{\lover}^{\lover-1}(\bfx)}{\bfx\in S}$ are 
pairwise-disjoint, hence the reconstruction algorithm 
of~\cref{lem:reconst-rf} may operate on all elements of~$S$ in 
parallel without interference.
\end{IEEEproof}
As a consequence of \cref{cor:reconst-rf-k}, $\rf_{\lover}(n,k)$ forms 
a multi-strand $(\lmin,\lover)$-trace code in $\cX_{n,k}$ (likewise, 
$\rf_{\lover}(n)$ in $\Sigma^n$).

Further, we note for $k=1$ that if $\liminf \lover/\log(n) > 1$, then 
\cite{EliGabMedYaa21} showed that $\rf_{\lover}(n)$ forms a 
rate~$1-o_n(1)$ code in $\Sigma^n$ with an efficient encoder/decoder 
pair. Before summarizing their results, we will require the following 
notation; let 
\begin{align*}
	\rll_s(n)\deq \mathset*{\bfu\in \Sigma^n}{\text{$\bfu$ has no 
	length-$s$ run of zeros}}.
\end{align*}
This is the well-understood \emph{run-length-limited} constraint (see, 
e.g., \cite[Sec.~1.2]{MarRotSie01}).

Then, from~\cite{EliGabMedYaa21} we have the following lemma.
\begin{lemma}\label{lem:elishco}
\begin{enumerate}
\item \cite[Sec.~IV]{EliGabMedYaa21}
There exists an efficient encoder/decoder pair into 
$\rf_{2\ceilenv*{\log(n)}+2}(n)$, requiring a single redundant symbol.

\item \label{it:elishco-red}\cite[Sec.~V]{EliGabMedYaa21}
There exists an efficient encoder/decoder pair into $\rf_{s'}(n)\cap 
\rll_{s''}(n)$, where 
\begin{align*}
	s' &\deq \ceilenv*{\log(n)} + 10\ceilenv*{\log\log(n)} + 10; \\
	s'' &\deq 4\ceilenv*{\log\log(n)} + 2.
\end{align*}
The required redundancy is 
\begin{align*}
	s'' + 1 + 
	\red\parenv*{\rll_{2\ceilenv*{\log\log(n)}}(n - s'' - 1)}.
\end{align*}
\end{enumerate}
\end{lemma}

Analysis of the asymptotic rate achieved by the encoder of 
\cref{lem:elishco}, \cref{it:elishco-red} is given in the following 
lemma.
\begin{lemma}\label{lem:rll-red}
There exist efficient encoders into $\rll_s(n)$ requiring 
$2\ceilenv*{2 n/2^s}$ redundant symbols for 
$q=2$~\cite[Sec.~III]{LevYaa19}, or $\ceilenv[\big]{\frac{q}{q-2} 
n/q^s}$ for $q>2$.
\end{lemma}
\begin{IEEEproof}
The claim for $q=2$ is proven in~\cite[Sec.~III]{LevYaa19}. Hence, we 
need only extend it when $q>2$, and to do so we rely on the concept of 
the encoder in~\cite[Alg.~1]{LevYaa19}. 
First, the information string~$\bfx\in\Sigma^m$ is divided into blocks 
of length~$N$ (where the last block is permitted to be shorter), to be 
determined later. Then, in each block:
\begin{enumerate}
    \item Append a $1$ to the block.
    \item \label{it:rll-replace}
    From left to right, search for zero-runs of length~$s$; if 
    one is encountered, remove it, and append the index of its 
    incidence to the block using $s$~symbols, such that the last 
    symbol is restricted not to be either $\bracenv*{0,1}$.
    \item Continue, until no further zero-runs of length $s$ exist.
\end{enumerate}

Note that this process concludes in finite time (since in each 
iteration of~\cref{it:rll-replace} it advances by at least~$s$ 
locations of the original block, and appended symbols contain no 
zero-run of length~$s$). Further, with the given restriction, 
$s$~symbols may index a total of $q^{s-1}(q-2)$ locations for the 
beginning of the zero $s$-substring. It is therefore required to set 
$N\deq q^{s-1}(q-2)+s-1$.

Also observe that a possible decoder can use the last symbol to 
indicate whether a zero-run of length~$s$ was removed and indexed 
(which it can then inject in the correct place, discarding the index), 
or if the process is concluded (in which case the suffix~`$1$' should 
also be discarded).

Next, since every encoded block ends with a nonzero symbol, these 
blocks can be concatenated without violating the constraint. Observe, 
then, that a single redundant symbol is added per block, hence the 
claimed overall redundancy.

Finally, note that both encoder and decoder operate in polynomial time 
in the input length.
\end{IEEEproof}
\cref{lem:rll-red} provides efficient encoders/decoders into 
$\rll_s(n)$; to complete the picture, we observe that their redundancy 
has asymptotically optimal order of magnitude; indeed, 
by~\cite[Lem.~3]{LevYaa19} we have that $\red(\rll_s(n))\geq 
\frac{\log(e)}{2} \parenv[\big]{1-\frac{1}{q}}^2 \frac{n-2s}{q^s}$.

Next, although the encoder of~\cref{it:elishco-red} 
of~\cref{lem:elishco} asymptotically achieves rate $1$, it is of 
interest to encode into $\rf_\ell(n)$ using less redundancy, for any 
$\ell < 2\log(n)$. We will show that the approach 
of~\cite[Sec.~V]{EliGabMedYaa21} can be generalized to this end. 

\begin{theorem}\label{cor:rf}
For integers $\ell(n), t$ satisfying 
\begin{align*}
	\ceilenv*{\log\log(n)} + 4 \leq t 
	\leq \floorenv*{\parenv*{\ell(n) - \ceilenv*{\log(n)}}/3}
\end{align*}
(for $q=2$, require 
$
\ceilenv*{\log\log(n)} + 5 \leq t \leq \floorenv*{\parenv*{\ell(n) - 
\ceilenv*{\log(n)}}/3}$) there exists an efficient encoder/decoder 
pair into $\rf_{\ell(n)}(n)\cap \rll_t(n)$, requiring at most $t+1+
\ceilenv[\big]{\frac{q^4}{q-2} n/q^t}$ redundant symbols (for $q=2$, this is $t+1+2\ceilenv*{16 n/2^t}$), i.e., rate 
$1-O_n\parenv*{\frac{t}{n} + q^{-t}}$.
\end{theorem}
\begin{IEEEproof}
The proof follows the steps of~\cite[Sec.~V]{EliGabMedYaa21}, with 
some amendments; where their arguments hold without change, we shall 
clearly cite the relevant proposition while reproducing its proof 
(when possible, we prioritize intuition over formality in our proof, 
without sacrificing rigour).
The rest of the proof is organized in stages, to improve readability.

\begin{enumerate}
\item 
In the first stage, an information string~$\bfx\in \Sigma^m$ is 
encoded into $\bfy\in \rll_{t-3}(n-t-1)$, where $m$~is determined by, 
e.g., \cref{lem:rll-red}.

\item 
Next, we wish to eliminate from $\bfy$ repeated substrings of 
length~$s\deq \ceilenv*{\log(n)} + t + 2$. 
The elimination stage requires an indexing function~$\bfh\colon [n]\to 
\rll_{t-3}(N)$ (i.e., an integer $N$ satisfying $\abs*{\rll_{t-3}(N)} 
\geq n$). 
By~\cref{lem:rll-red} an explicit function exists if 
$\ceilenv[\big]{\frac{q}{q-2} N/q^{t-3}}\leq N-\log(n)$ (for $q=2$ 
that is $2\ceilenv*{2 N/2^{t-3}}\leq N-\log(n)$), or equivalently 
$\parenv[\big]{1-\frac{q^{4-t}}{q-2}} N\geq \ceilenv*{\log(n)}$. With 
the assumed lower bound on~$t$, this requirement is satisfied by
$N\deq \ceilenv*{\log(n)} + 1$, for sufficiently large~$n$.

\item 
In the elimination stage (based on~\cite[Alg.~3]{EliGabMedYaa21}), 
$\bfy$ is processed from left to right; whenever $j>i$ are found such 
that $\bfy_{j+[s]} = \bfy_{i+[s]}$ (and again, $j$ is minimal 
satisfying this requirement), the segment $\bfy_{j+[s]}$ is deleted 
and replaced with 
\begin{align*}
	1\; 0^{t-3}\; 1\circ \bfh(i)\circ 1, 
\end{align*}
where we consider $1 0^{t-3} 1$ to be a marker, indicating the 
replaced segment (based on the first step, this marker does not appear 
elsewhere in $\bfy$). 
Based on the fact that any elimination reduces the string length 
by~$1$, this stage is concluded in $O(n^2)$ steps. We denote the 
resulting string by~$\bfw$, of length~$n'$ (for some $n'\leq n-t-1$, 
depending on how many eliminations were performed). Trivially, the 
only instances of~$0^{t-3}$ in~$\bfw$ are the markers used in replaced 
substrings (and $\bfw\in \rll_{t-2}(n')$). By following the same 
approach as in~\cite[Lem.~19]{EliGabMedYaa21} (which in turn was based 
on~\cite[Cla.~10]{GabMil19}) one observes that $\bfw\in \rf_s(n')$ and 
that the process can be reversed; the former is trivial since the 
process only terminates when no repeated sequences remain. The latter 
is done by decoding from right to left, where replaced substrings are 
identified by the presence of markers, and the eliminated substrings 
are restored based on~$\bfh(i)$. 
To prove this is possible, one needs only show that after 
any iteration of the process, the right-most instance of a marker is 
the one injected in that iteration. Indeed, since the process scans 
for~$j$ from left to right, if $j$ is the location identified (i.e., 
$\bfy_{j+[s]}$ was replaced) in the last iteration, and $j'$ in the 
iteration before that, then by necessity $j\geq j'-s+1$; clearly, 
then, if $j<j'$ then the marker injected at location~$j'$ was 
overwritten in the last iteration. I.e., the marker injected at any 
iteration either overwrites the last injected marker, or appears in 
the replaced string to its right.

\item 
The process is concluded in an expansion stage, meant to output 
strings of length~$n$ from which $\bfw$ (hence also $\bfx$) can be 
decoded. For that purpose, an arbitrary string $\bfv\in \rf_s(n) 
\cap \rll_{t-2}(n)$ is generated in a fashion to be described below, 
and interleaved with $1\; 0^{t-2}\; 1$ marker-segments after every 
$s$~positions; i.e., if $\bfv = \bfv_0\circ \bfv_1\circ \cdots\circ 
\bfv_{\ceilenv*{n/s}-1}$, where $\abs*{\bfv_i} = s$ for all $i\in 
[\floorenv*{n/s}]$ and $\abs*{\bfv_{\ceilenv*{n/s}-1}}\leq s$, then 
\begin{align*}
	\bfw'\deq \bfv_0\circ 1\; 0^{t-2}\; 1\circ \bfv_1\circ 
	1\; 0^{t-2}\; 1\circ \cdots\circ \bfv_{\ceilenv*{n/s}-1}.
\end{align*}
Clearly, $\bfw'\in \rf_{s+t}(n'')\cap \rll_{t-1}(n'')$, where $n''\deq 
\abs*{\bfw'}\geq n$. It is straightforward that the only instances 
of~$0^{t-2}$ in~$\bfw'$ are the markers interleaved into it. Based on 
these observations, it is proven similarly 
to~\cite[Lem.~23]{EliGabMedYaa21} that 
\begin{align*}
	\hat{\bfw}\deq \bfw \circ 1\; 0^{t-1}\; 1\circ \bfw' 
\end{align*}
is $(s+2t-2) = (\ceilenv*{\log(n)} + 3t)$-repeat-free and 
$t$-run-length-limited; this is done by observing that any substring 
of this length of~$\hat{\bfw}$ contains markers (potentially unless it 
is a substring of~$\bfw$, in which case the absence of markers 
indicates that fact), their length ($(t-i)$ for $i\in 
\bracenv*{1,2,3}$) indicates which portion of of~$\hat{\bfw}$ it is 
taken from, and if it does not cover the unique instance of~$0^{t-1}$ 
in~$\hat{\bfw}$ then it contains $s$~consecutive symbols of either 
$\bfw$ or $\bfv$, hence is unique.

Observe that by the upper bound on~$t$, $\hat{\bfw}$ is 
$\ell(n)$-repeat-free. Also, the $n$-prefix of~$\hat{\bfw}$ contains 
$\bfw\circ 1\; 0^{t-1}\; 1$, hence $\bfw$ can uniquely be extracted 
from it. That prefix is therefore output as the encoded information.

\item 
Finally, it remains to describe how any arbitrary $\bfv\in \rf_s(n) 
\cap \rll_{t-2}(n)$ might be generated (a single example suffices). 
To achieve this, any total order on $\Sigma$ is chosen where $0$ is 
the minimum, and $1$ the maximum.

For a string~$\bfu\in\Sigma^s$, let its \emph{necklace} be the 
lexicographic least cyclic rotation of~$\bfu$, and its 
\emph{periodic-reduction} be the minimum period of~$\bfu$. It was 
shown in~\cite{FreMai78} that concatenating in lexicographic 
order periodic-reductions of necklaces of length~$s$ produces a 
de~Bruijn sequence~$\bfb\in \Sigma^{q^s}$ (in fact, this is the 
lexicographically least de~Bruijn sequence of that length).

The last instance of $0^t$ in~$\bfb$ is in the necklace $0^t\; 
1^{s-t}$ (since only the ``$0$'' necklace ends with $0$, and the 
longest zero-run in any necklace appears at its beginning). Hence, 
letting $i\in [q^s]$ be the unique location such that $\bfb_{i+[s]} = 
0^t\; 1^{s-t}$, we let $\bfv\deq \bfb_{q^s-n+[n]}$ and to conclude we 
need only show that $q^s-n>i$.

This was done in~\cite[Lem.~20]{EliGabMedYaa21} in case that $s$ is 
prime, and we generalize for all~$s$; we do so by counting 
$\abs*{\mathset*{\bfb_{j+[s]}}{i<j\leq q^s-s}} = q^s-s-i$. By the 
proof of~\cite[Th.~4]{FreMai78} every $\bfu\in \Sigma^s$ appears 
in~$\bfb$ in a location intersecting the appearance of the 
periodic-reduction of its necklace, potentially unless $u_0 = 1$.

Observe that it is sufficient that $\bfu\in \rll_t(s)$ satisfies 
$u_{s-1}\neq 0$ for its necklace to be greater than $0^t\; 1^{s-t}$; 
therefore for all $\bfu\in \rll_t(s-2)$ and $u', u''\in \Sigma$ 
satisfying $u'\neq 1$, $u''\neq 0$ there exists $j>i$ such that 
$\bfb_{j+[s]} = u'\circ \bfu\circ u''$. It follows that 
\begin{align*}
	q^s-s-i &= \abs*{\mathset*{\bfb_{j+[s]}}{i<j\leq q^s-s}} \\
	&\geq (q-1)^2 \abs*{\rll_t(s-2)} \\
	&\overset{(*)}{\geq} (q-1)^2 q^{(s-2) \parenv*{1-2 q^{-t}}} \\
	&\geq (q-1)^2 n^{(1+t/\log(n)) \parenv*{1-2 q^{-t}}},
\end{align*}
where $(*)$ is justified by~\cite[Lem.~3]{LevYaa19} for sufficiently 
large~$t$. Since $t > \ceilenv*{\log\log(n)}$, for sufficiently 
large~$n$ we have $(1+t/\log(n)) \parenv*{1-2 q^{-t}} > 1$, as 
required.
\end{enumerate}

Finally, redundancy of this construction is $t+1$, plus the redundancy 
of encoding into $\rll_{t-3}(n-t-1)$; \cref{lem:rll-red} now concludes 
the proof.
\end{IEEEproof}

In summary, we have the following corollary:
\begin{corollary}
By~\cref{lem:reconst-rf}, $\rf_{\lover}(n)$ forms an $(\lmin, 
\lover)$-trace code in $\Sigma^n$, which by~\cref{cor:rf} has 
$1-o_n(1)$ rate whenever $\lover \geq \ceilenv*{\log(n)} + 
3 \ceilenv*{\log\log(n)} + 12$.
\end{corollary}
In the sequel, we therefore focus on the complement, unsolved case of 
$\limsup \lover/\log(n) \leq 1$.

\section{Bounds}\label{sec:partial-single}

In this section we demonstrate an upper bound on the achievable 
asymptotic rate of $(\lmin, \lover)$-trace codes.

\begin{lemma}\label{lem:code-size}
Any multi-strand $(\lmin, \lover)$-trace code~$\cC\subseteq \cX_{n,k}$ 
satisfies 
\begin{align*}
	\abs*{\cC} 
	&\leq \mybinom{k \ceilenv[\big]{\frac{n-\lover}{\lmin-\lover}} + 
	q^{\lmin}}{q^{\lmin}}.
\end{align*}
\end{lemma}
\begin{IEEEproof}
Since $\cL_{\lmin}^{\lover}(\bfx)\in \cT_{\lmin}^{\lover}(\bfx)$ for 
all $\bfx\in \Sigma^n$, we have 
\begin{align*}
	\abs*{\cC} &\leq 
	\abs*{\mathset*{\cL_{\lmin}^{\lover}(S)}{S\in \cX_{n,k}}}.
\end{align*}

Similarly to the argument used in~\cite{ChaChrEzeKia17}, we count the 
incidences of each possible $\bfu\in \Sigma^{\lmin}$ in 
$\cL_{\lmin}^{\lover}(S)$, resulting in $f_S\colon \Sigma^{\lmin}\to 
\N$ (dubbed a \emph{profile-vector} in~\cite{ChaChrEzeKia17}). Observe 
that $\sum_{\bfu\in \Sigma^{\lmin}} f_S(\bfu) 
= k \parenv*{1+\ceilenv[\big]{\frac{n-\lmin}{\lmin-\lover}}} 
= k \ceilenv[\big]{\frac{n-\lover}{\lmin-\lover}}$; thus, we have an 
embedding of $\mathset*{\cL_{\lmin}^{\lover}(S)}{S\in \cX_{n,k}}$ into 
\begin{align*}
	\mathset*{\bff\in \N^{q^{\lmin}}}{\sum_{\mathclap{i\in 
	[q^{\lmin}]}} \bff_i = k\ceilenv*{\frac{n-\lover}{\lmin-\lover}}},
\end{align*}
and therefore 
\begin{align*}
	\abs*{\cC} 
	&\leq \mybinom{k \ceilenv[\big]{\frac{n-\lover}{\lmin-\lover}} + 
	q^{\lmin} - 1}{q^{\lmin} - 1},
\end{align*}
which concludes the proof.
\end{IEEEproof}

\begin{lemma}\label{lem:over-lin-rate}
For $k=1$, if $\lmin = a \log(n) + O_n(1)$ and $\lover = \gamma \lmin 
+ O_n(1)$, for some $a > 1$ and $0 \leq \gamma\leq \frac{1}{a}$, then 
any $(\lmin, \lover)$-trace code $\cC\subseteq \Sigma^n$ satisfies 
\begin{align*}
	R(\cC) 
	\leq \frac{1-1/a}{1-\gamma} + 
	O\parenv*{\frac{\log\log(n)}{\log(n)}}.
\end{align*}
(Note that $\gamma$ is a linear scaling of the required overlap 
between consecutive segments, in proportion to their required minimum 
length. We scale that minimum length linearly with~$\log(n)$ (where 
the constant~$a$ indicates the ratio), a decision informed by the 
statement of this lemma, and the succeeding corollary. 
Finally, observe that in this notation, $\frac{1 - 1/a}{1-\gamma}\leq 
1$ if and only if $\lim \lover/\log(n) = \gamma a \leq 1$.)
\end{lemma}
\begin{IEEEproof}
From the known bound $u! > (u/e)^u$ we observe for all $v\geq u > 0$ 
that 
\begin{align*}
	\log\binom{u+v}{u} &\leq \log\frac{(u+v)^u}{u!} 
	< \log\parenv*{\parenv*{\tfrac{e}{u}(u+v)}^u} \\
	&= u \parenv*{\log(e) + \log\parenv*{1+\tfrac{v}{u}}} \\
	&= u \parenv*{\log(e) + \log\parenv*{\tfrac{u}{v}+1} + 
	\log\parenv*{\tfrac{v}{u}}} \\
	&< u \parenv*{2 \log(e) + \log\parenv*{\tfrac{v}{u}}}, 
\end{align*}
where the last inequality holds since $\tfrac{u}{v}+1\leq 2 < e$.

Letting $v\deq q^{\lmin}$ and $u\deq 
\ceilenv[\big]{\frac{n-\lover}{\lmin-\lover}} < v$, we observe that 
$\log(\frac{v}{u}) = O(\log(n))$ and $\log(u) \geq \log(n-\lover) - 
\log\log(n) + O(1)$; observing 
\begin{align*}
	\log(n-\lover) &= \log(n) + \log\parenv*{1-\tfrac{\lover}{n}} \\
	&\geq \log(n) - \frac{\log(e) \lover}{n - \lover} \\
	&= \log(n) - O\parenv*{\frac{\log(n)}{n}}, 
\end{align*}
where we used $\ln(1-x)\geq \frac{-x}{1-x}$, we summarize $\log(u) 
\geq \log(n) - \log\log(n) + O(1)$.

Next, by~\cref{lem:code-size} $\abs*{\cC}\leq \binom{u+v}{u}$, hence 
we have 
\begin{IEEEeqnarray*}{+rCl+x*}
	\log\abs*{\cC} 
	&\leq& \parenv*{\frac{n-\lover}{\lmin-\lover} + 1} 
	\parenv*{\log\parenv*{\tfrac{v}{u}} + 2 \log(e)} \\
	&=& \frac{n \parenv*{\log\parenv*{\tfrac{v}{u}} + 
	2 \log(e)}}{\lmin-\lover} + O\parenv*{\log\parenv*{\tfrac{v}{u}}} \\
	&=& n \frac{\log(v)-\log(u)}{\lmin-\lover} + 
	O\parenv*{\frac{n}{\log(n)}} + O\parenv*{\log(n)} \\
	&=& n \parenv*{\frac{\log(v)-\log(u)}{\lmin-\lover} + 
	O\parenv*{\frac{1}{\log(n)}}} \\
	&\leq& n \parenv*{\frac{\lmin - \log(n)}{\lmin-\lover} 
	+ O\parenv*{\frac{\log\log(n)}{\log(n)}}} \\
	&=& n \parenv*{\frac{1-1/a}{1-\gamma} 
	+ O\parenv*{\frac{\log\log(n)}{\log(n)}}}.
	\\[-\normalbaselineskip] &&&\IEEEQEDhere
\end{IEEEeqnarray*}
\end{IEEEproof}

In particular, \cref{lem:over-lin-rate} implies the following lower 
bound on $\lmin$ for the existence of codes with asymptotically 
non-vanishing rates.
\begin{corollary}
Take $(\lmin^{(n)})_{n>0}, (\lover^{(n)})_{n>0}$, and let $\cC^n 
\subseteq \Sigma^n$ be $(\lmin^{(n)}, \lover^{(n)})$-trace codes. 
If $\limsup_n \lmin^n/\log(n)\leq 1$, then $R(\cC^n)=o_n(1)$.
\end{corollary}
\begin{IEEEproof}
Since $\cC^n$ are also $(\lmin'{}^{(n)}, 0)$-trace codes for 
$\lmin'{}^{(n)}\geq \lmin^{(n)}$, it follows from 
\cref{lem:over-lin-rate} that $R(\cC^n)\leq \frac{1-1/a}{1-0}+o(1)$ 
for all $a>1$, hence the claim follows.
\end{IEEEproof}

\section{A Construction of Trace Codes}
\label{sec:partial-single-encode}

In this section we present an efficient encoder for $(\lmin, 
\lover)$-trace codes (i.e., in the case $k=1$), achieving 
asymptotically optimal rate, for the case $\limsup \lover/\log(n) \leq 
1$ (complementing the results of~\cref{sec:def}). Throughout the 
section, we let 
\begin{align}\label{eq:lmin-lover}
	\lmin &\deq \ceilenv*{a \log(n)}; 
	\nonumber \\
	\lover &\deq \ceilenv*{\gamma \lmin}, 
\end{align}
for some $a>1$ and $0 < \gamma \leq 1/a$. 
Further, we let $f$ be any integer function satisfying $f(n) = 
o(\log(n))$ and $f(n) \geq \log\log(n) + 4$, and finally 
\begin{align}\label{eq:I-def}
	I &\deq \ceilenv*{\frac{1-\gamma a}{1-\gamma} \log(n) 
	+ (\log(n))^{0.5+\epsilon}}, 
\end{align}
for some small $\epsilon>0$. 
In our construction, $I$ is the number of symbols dedicated to 
(unencoded-)indices, which are then partitioned into length-$f(n)$ 
fragments, as described below. 
When analyzing the redundancy of our construction, we shall optimize 
it by a proper choice of~$f(n)$, in~\cref{thm:single-red}.

The main idea of the construction presented below of an $(\lmin, 
\lover)$-trace code~$\code{cnst:overlap}(n)$ is to encode an 
information string~$\bfx$ into $\parenv*{\bfz_i}_{i\in [q^I]}$ so that 
the following two properties are satisfied:
\begin{enumerate*}[label=(\roman*)]
\item the index~$i$ can be decoded from any $\lmin$-substring 
of~$\bfz_i$; and 
\item the string~$\bfz_i$ can be uniquely reconstructed from an 
$(\lmin, \lover)$-trace of~$\bfz_i$.
\end{enumerate*}
This is performed by interleaving segments of indices in 
appropriate locations in the encoded strings. 
Then, we let
\begin{align*}
	\enc[cnst:overlap](\bfx)\deq 
	\bfz = \bfz_0\circ \cdots\circ \bfz_{q^I-1}\in 
	\code{cnst:overlap}(n).
\end{align*}

Before presenting the construction, we describe the method 
of index-generation.
\begin{definition}\label{def:index}
Let $\parenv*{\bfc_i}_{i\in[q^I]}$, $\bfc_i\in\Sigma^I$ be indices in 
ascending lexicographic order. We encode each $\bfc_i$ independently 
as follows (see~\cref{fig:index_gen}). 
Denoting $F\deq \ceilenv*{I/f(n)}$, we partition $\bfc_i$ into~$F$ 
non-overlapping segments of equal lengths 
$\bracenv[\big]{\bfc_i^{(h)}}_{h\in [F]}$; here and in the sequel, we 
say a string is partitioned into non-overlapping segments of equal 
lengths if 
$\bfc_i^{(0)}\circ \bfc_i^{(1)}\circ \cdots\circ \bfc_i^{(F-1)} = 
\bfc_i$ and 
\begin{align*}
	\abs[\big]{\bfc_i^{(h)}} = \begin{cases}
		\ceilenv*{I/F}, & h<I\bmod F; \\
		\floorenv*{I/F}, & \text{otherwise}.
	\end{cases}
\end{align*}

Observe that $\abs[\big]{\bfc_i^{(h)}}\leq f(n)$ for all~$h\in [F]$. 
We then denote $\bfc'_i{}^{(h)}\deq 1\circ \bfc_i^{(h)}\circ 1$. We 
refer to~$\bfc_i$ (or simply~$i$) as an \emph{index} in the 
construction, and to~$\bracenv*{\bfc'_i{}^{(h)}}_{h\in [F]}$ as 
segments of an \emph{encoded index}.

\begin{figure}[t]{}%
\centering
\psfrag{parity symbol}{parity symbol}
\includegraphics[width=0.95\columnwidth]{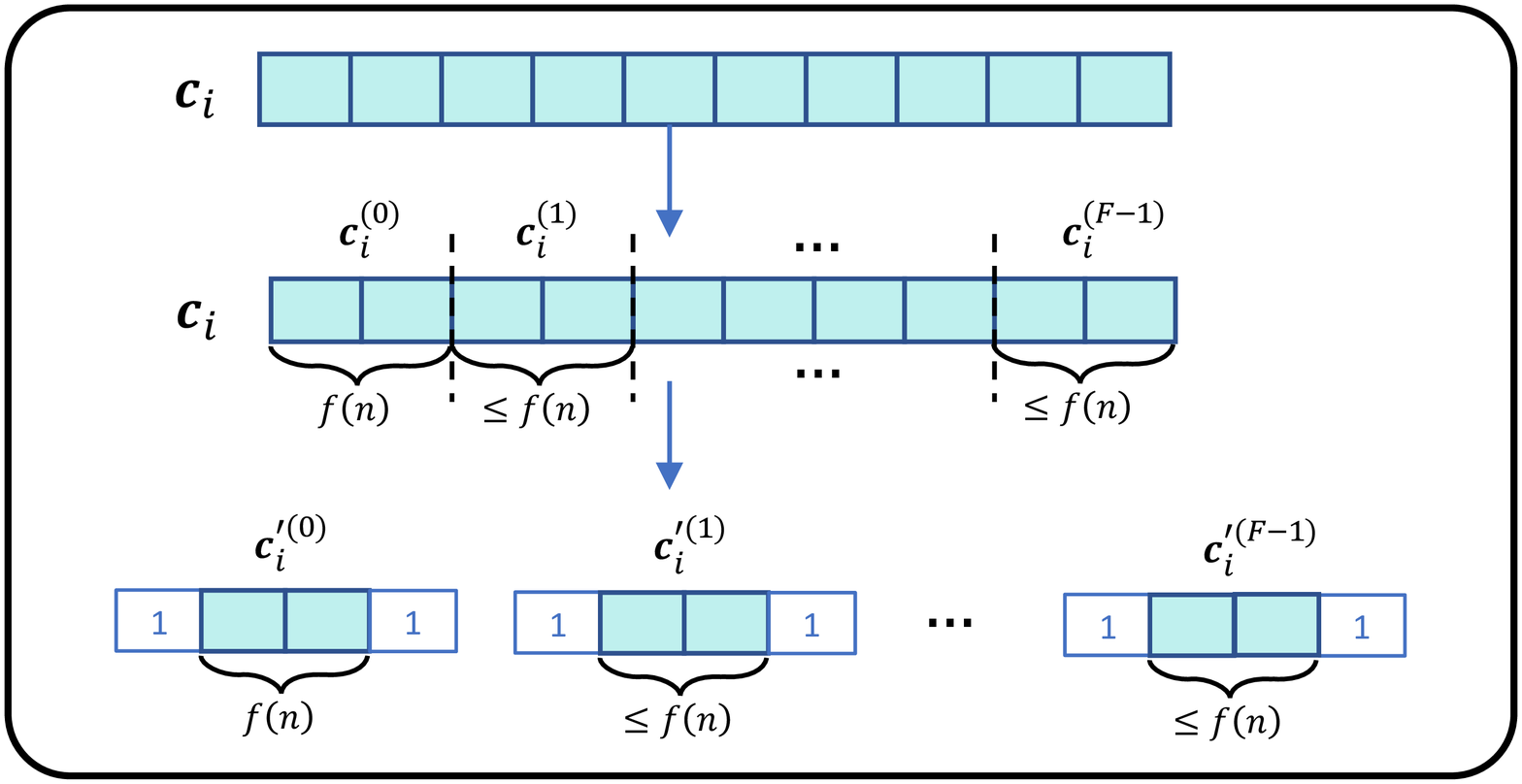}%
\caption{Index generation. Each index $\bfc_i$ is first partitioned 
into $F+1$ non-overlapping segments of length $f(n)$. Then, each of 
the segments is concatenated with a single $1$ in each edge.   
\label{fig:index_gen}}
\end{figure}
\end{definition}

Further, for $N\leq n$ to be defined later, and $\ell > \ceilenv*{\log(N)} + 
3 f(n)$, we denote the encoder of~\cref{cor:rf} 
\begin{align*}
	E^{\rf}_{N,\ell}\colon \Sigma^{m(N)}\to \rf_\ell\parenv*{N}
	\cap \rll_{f(n)+1}\parenv*{N}.
\end{align*}
Here, 
\begin{align}\label{eq:m}
	m(N) &\deq N - \red(E^{\rf}_{N,\ell}) 	\nonumber \\
	&\geq N \parenv*{1 - \frac{f(n)+2}{N} - \frac{q^3}{q-2}q^{-f(n)}}.
\end{align}
(For $q=2$, $m(N)\geq N \parenv*{1 - \frac{f(n)+4}{N} - 2^{4-t}}$.)

\begin{construction}\label{cnst:overlap}
The encoding into~$\bfz_i$, for all $i\in [q^I]$, is performed as 
follows (see~\cref{fig:x_i encoding}). We denote 
\begin{align}\label{eq:r-def}
	r &\deq I + 2F + f(n) + 4 \nonumber \\
	&= I + \frac{2I}{f(n)} + f(n) + O(1), 
\end{align}
then define 
\begin{align}\label{eq:ell-def}
	\ell \deq \ceilenv[\bigg]{\frac{\lover-2f(n)-6}{1 + 
	(f(n)+2)\big/\floorenv[\big]{\frac{\lmin-r}{F}}}}.
\end{align}
(see \cref{lem:suff-pre-pair,lem:consec}, respectively, for 
the reason for these definitions). Also, for all $i\in [q^I]$ 
\begin{align}\label{eq:N-def}
	N_i\deq \begin{cases}
		\ceilenv*{q^{-I} n} - \ceilenv*{n/(q^I \lmin)} r, & 
		i < n\bmod q^I; \\
		\floorenv*{q^{-I} n} - \ceilenv*{n/(q^I \lmin)} r, & 
		\text{otherwise}.
	\end{cases}
\end{align}
Now, for all $i\in [q^I]$ define $\bfy_i\deq 
E^{\rf}_{N_i,\ell}(\bfx_i)\in \Sigma^{N_i}$, where $\bfx_i\in 
\Sigma^{m(N_i)}$ and 
\begin{align*}
	\bfx = \bfx_0\circ \bfx_1\circ \cdots\circ \bfx_{q^I-1} 
\end{align*}
is an arbitrary information string (see the proof 
of~\cref{thm:single-red} for a choice of~$f(n)$ satisfying the 
conditions of~\cref{cor:rf}, hence assuring the existence of 
$E^{\rf}_{N_i,\ell}$).

Next, for all $i\in [q^I]$ 
\begin{enumerate}
\item
Partition $\bfy_i$ into $\ceilenv*{n/(q^I \lmin)}$ non-overlapping 
segments of equal length 
\begin{align*}
	\bfy_i 	= \bfy_{i,0}\circ \bfy_{i,1}\circ \cdots\circ 
	\bfy_{i,\ceilenv*{n/(q^I \lmin)}-1}.
\end{align*}

\item\label{step:index-injection}
For all $j\in [\ceilenv*{n/(q^I \lmin)}]$:
\begin{enumerate}
	\item
	Partition each $\bfy_{i,j}$ into $F$ non-overlapping segments of 
	equal lengths 
	\begin{align*}
		\bfy_{i,j} = \bfy_{i,j}^{(0)}\circ \bfy_{i,j}^{(1)}\circ 
		\cdots\circ \bfy_{i,j}^{(F-1)}.
	\end{align*}
	
	\item
	Combine $\mathset[\big]{\bfy_{i,j}^{(h)}}{h\in [F]}$ with 
	segments of the encoded index~$i$, as follows. Define for all 
	$h\in [F]$ 
	\begin{align*}
		\bfz_{i,j}^{(h)} \deq \bfy_{i,j}^{(h)}\circ \bfc'_i{}^{(h)}, 
	\end{align*}
	then 
	\begin{align*}
		\bfz_{i,j}\deq \begin{cases}
		1\; 0^{f(n)+1}\; 1\; 1\circ \bfz_{i,j}^{(0)}\circ \cdots
		\circ \bfz_{i,j}^{(F-1)}, & j=0; \\
		1\; 0^{f(n)+1}\; 0\; 1\circ \bfz_{i,j}^{(0)}\circ \cdots
		\circ \bfz_{i,j}^{(F-1)}, & j>0
		\end{cases}
	\end{align*}
	(we refer to the substrings $1 0^{f(n)+1} 1 1, 1 0^{f(n)+1} 0 1$ 
	as \emph{synchronization markers}).
\end{enumerate}

\item
Concatenate 
\begin{align*}
	\bfz_i &\deq \bfz_{i,0}\circ \cdots\circ 
	\bfz_{i,\ceilenv*{n/(q^I \lmin)}-1}.
\end{align*}
\end{enumerate}

\begin{figure}[t]{}%
\centering
\includegraphics[width=0.95\columnwidth]{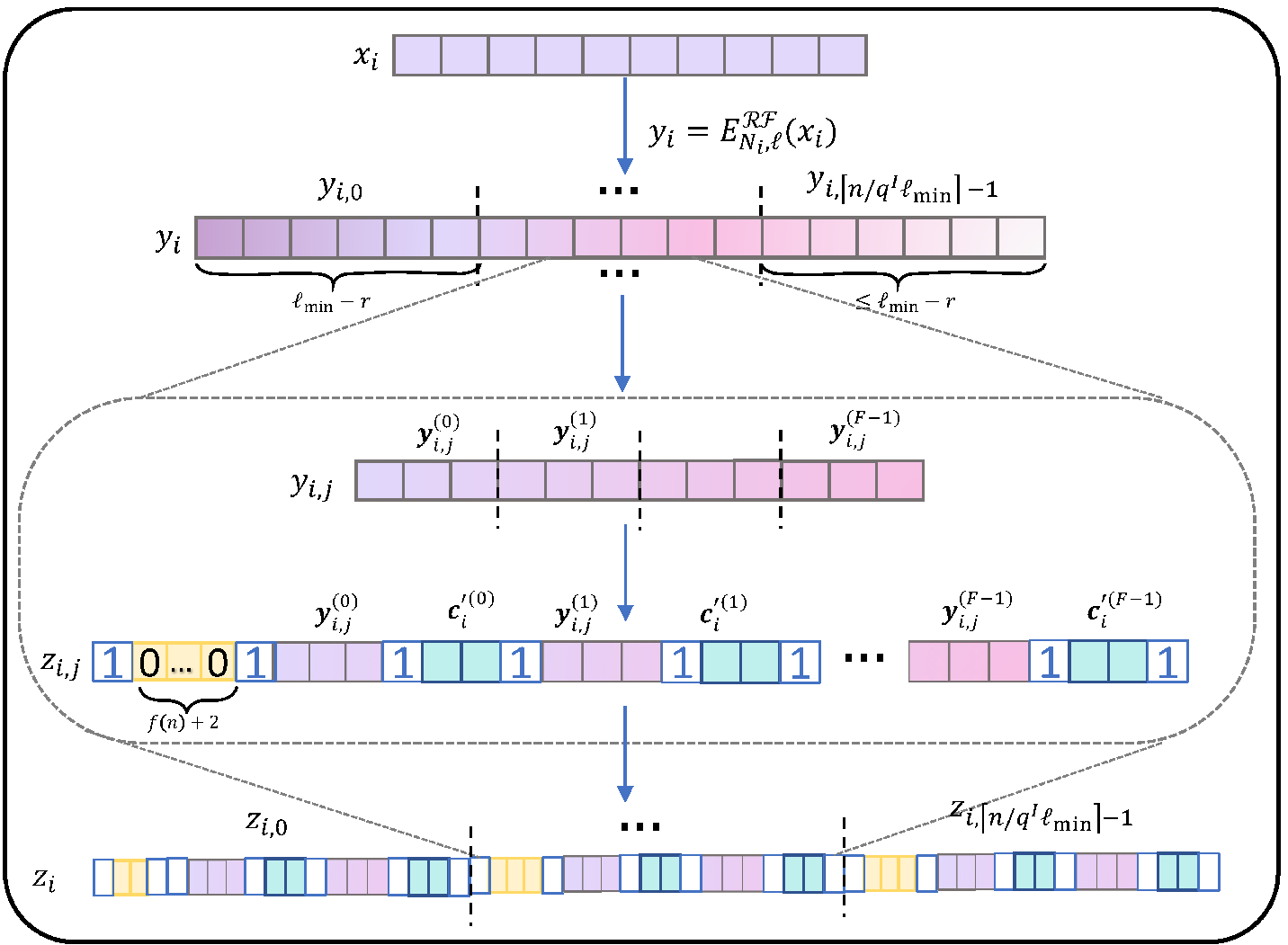}%
\caption[Encoding]{Encoding $\bfx_i$ into $\bfz_i$, as detailed in 
\cref{cnst:overlap}.
\label{fig:x_i encoding}}
\end{figure}

\vspace*{-1.5\normalbaselineskip}
\end{construction}

First, we prove the correctness of \cref{cnst:overlap}. We begin 
with two technical lemmas which are key to the proof of correctness in 
\cref{thm:overlap-correct}.
\begin{lemma}\label{lem:suff-pre-pair}
Every $\lmin$-substring $\bfu$ of $\bfz \in \code{cnst:overlap}(n)$ 
contains as subsequences at least an $(I-\mu)$-suffix of an 
index~$\bfc_i$ (see \cref{def:index}), and an $\mu$-prefix of either 
$\bfc_i$ or $\bfc_{i+1}$, for some $i\in [q^I]$ and $\mu\in [I]$, in 
identifiable locations.
\end{lemma}
\begin{IEEEproof}
Note that 
\begin{align*}
	\lmin - q^I \lmin^2/n &\leq \frac{\lmin}{1 + q^I \lmin/n} 
	= \frac{n/q^I}{n/(q^I \lmin)+1} \\
	&\leq \frac{n/q^I}{\ceilenv*{n/(q^I \lmin)}} 
	\leq \frac{\ceilenv{n/q^I}}{\ceilenv*{n/(q^I \lmin)}} 
	\leq \lmin
\end{align*}
Observing from \cref{eq:I-def} that $q^I \lmin^2 = o(n)$, and by 
subtracting~$r$ from the above inequality, it holds from 
\cref{eq:N-def} for sufficiently large~$n$ and all $i\in [q^I]$ that 
$\lmin-r-1 \leq N_i/\ceilenv*{n/(q^I \lmin)}\leq \lmin-r$. Hence also 
for all $j\in [\ceilenv*{n/(q^I \lmin)}]$ it holds that 
\begin{align}\label{eq:y_ij}
	\lmin-r-1 \leq \abs*{\bfy_{i,j}} \leq \lmin-r.
\end{align}
By~\cref{step:index-injection} of~\cref{cnst:overlap} it follows that 
$\abs*{\bfz_{i,j}} = \abs{\bfy_{i,j}}+r\in \bracenv*{\lmin,\lmin-1}$.

Next, observe that instances of synchronization markers only appear in 
$\bfz$ at the beginning of $\bracenv*{\bfz_{i,j}}_{i,j}$. From the 
last paragraph, either $\bfu$ contains a complete synchronization 
marker as substring, or it contains a suffix-prefix pair whose 
concatenation is an instance of a synchronization marker; in both 
cases, the exact locations in which symbols of the indices 
$\bracenv[\big]{\bfc'_i{}^{(h)}}$ appear can be determined. 
Extracting $\bracenv[\big]{\bfc_i{}^{(h)}}$, these contain a suffix of 
$\bfc_i$ and a prefix of either $\bfc_i, \bfc_{i+1}$ (depending on 
whether $\bfu$ is a substring of $\bfz_i$ for some~$i$) whose combined 
lengths is $I$, again since for all~$i,j$, $\abs*{\bfz_{i,j}}\leq 
\lmin$ and $\bfz_{i,j}$ contains all symbols of $\bfc_i$. 
Taking $\mu\in [I]$ to be the length of the prefix ($\mu = 0$ 
indicates the possibility that all symbols of the same index appear 
in~$\bfu$) concludes the proof.
\end{IEEEproof}

\begin{lemma}\label{lem:consec}
Every $\lover$-substring $\bfv$ of $\bfz\in \code{cnst:overlap}(n)$ 
contains at least~$\ell$ consecutive symbols of $\bfy\deq \bfy_0\circ 
\cdots\circ \bfy_{q^I-1}$ (see \cref{eq:ell-def}).
\end{lemma}
\begin{IEEEproof}
At worst, $\bfv$ either begins or ends with a complete instance of a 
synchronization marker; hence the remaining $\lover-f(n)-4$ symbols 
are sampled from $\bracenv[\big]{\bfz_{i,j}^{(h)}}$, and again, at 
worst end with a complete segment of an encoded index. Since from 
\cref{def:index} $\abs[\big]{\bfc'_i{}^{(h)}}\leq f(n)+2$ and 
by~\cref{eq:y_ij} $\abs[\big]{\bfy_{i,j}^{(h)}}\geq 
\floorenv*{\frac{\lmin-r}{F}}$ for all $i,j,h$, $\bfv$ contains at 
least 
\begin{align*}
	\ceilenv[\bigg]{\frac{\lover-2f(n)-6}{1 + 
	(f(n)+2)\big/\floorenv[\big]{\frac{\lmin-r}{F}}}} = \ell 
\end{align*}
consecutive symbols of $\bfy$.
\end{IEEEproof}

Combining both lemmas, we have the following theorem.
\begin{theorem}\label{thm:overlap-correct}
For all admissible values of $n$, the code $\code{cnst:overlap}(n)$ is 
an $(\lmin, \lover)$-trace code.
\end{theorem}
\begin{IEEEproof}
Take $\bfz\in \code{cnst:overlap}(n)$ and let $T\in 
\cT_{\lmin}^{\lover}(\bfz)$, i.e., any $(\lmin, \lover)$-trace 
of~$\bfz$.

For $\bfu\in T$, we extract the $(I-\mu)$-suffix of $\bfc_i$, and an 
$\mu$-prefix of either $\bfc_i$ or $\bfc_{i+1}$, for some $i$, 
guaranteed by \cref{lem:suff-pre-pair}. 
Observe that if this prefix belongs to $\bfc_{i+1}$, then $\bfu$ 
also contains a complete synchronization marker $1 0^f(n) 1 1$ (the 
instance appearing as prefix of $\bfz_{i+1}$), hence these two cases 
may be distinguished. Further, note that the $\mu$-prefix of 
$\bfc_{i+1}$ equals the $\mu$-prefix of $\bfc_i$, unless every symbol 
of the $(I-\mu)$-suffix of $\bfc_i$ is $(q-1)$, in which case it is 
the $q$-ary expansion of the successor natural number to that prefix. 
In both cases, one can correctly deduce that the location of $\bfu$ in 
$\bfz$ begins in the segment $\bfz_i$. It is therefore possible to 
partition $T$ by index~$i$ (corresponding to the starting location of 
each substring).

For each substring $\bfu$ of index~$i$, intersecting both $\bfy_i, 
\bfy_{i+1}$, $\bfu$ must contain a complete synchronization marker 
$1 0^f(n) 1 1$ (the instance appearing as prefix of $\bfz_{i+1}$); 
hence its location in $\bfu$ implies the exact location of $\bfu$ in 
$\bfz$. 
For all other substrings of index~$i$, it holds by \cref{lem:consec}, 
and since each $\bfy_i$ is $\ell$-repeat-free, that there exist a 
unique way to concatenate these substrings (excluding overlap) as 
shown in \cref{lem:reconst-rf}.

Finally, once $\bfz$ is reconstructed we may extract 
$\bracenv*{\bfy_i}_{i\in [q^I]}$, then decode 
$\bracenv*{\bfx_i}_{i\in [q^I]}$ with the decoder of 
$E^{\rf}_{N,\ell}$.
\end{IEEEproof}

Next, we analyze $R\parenv*{\code{cnst:overlap}(n)}$. 
First, we require a simplified (asymptotic) expression 
for~$\ell$, used in~\cref{cnst:overlap} for repeat-free encoding, 
which we derive in the next lemma.
\begin{lemma}\label{lem:overlap-prop}
Denoting $\lambda\deq 1-\frac{I}{\lmin}$, we have 
\begin{align*}
	\ell = \lambda \lover - O\parenv*{f(n) + \frac{\log(n)}{f(n)}}.
\end{align*}
\end{lemma}
\begin{IEEEproof}
Recall the definition $\ell = \ceilenv[\Big]{\frac{\lover-2f(n)-6}{1 + 
(f(n)+2)\big/\floorenv[\big]{\frac{\lmin-r}{F}}}}$ 
in~\cref{eq:ell-def}, where $F = \ceilenv*{I/f(n)}$ and $r$~is defined 
in~\cref{eq:r-def}. We begin by observing 
\begin{align*}
	\frac{f(n)+2}{\floorenv[\big]{\frac{\lmin-r}{F}}} 
	&= \frac{F (f(n)+2)}{\lmin - r - O(F)} \\
	&= \frac{I + O\parenv*{f(n)}}{\lmin - I - 	\frac{2 I}{f(n)} - 
	O\parenv*{f(n) + \frac{\log(n)}{f(n)}}} \\
	&= \frac{I}{\lmin - I}\cdot \frac{1 + 
	O\parenv*{\frac{f(n)}{\log(n)}}}{1 - O\parenv*{\frac{1}{f(n)} + 
	\frac{f(n)}{\log(n)}}} \\
	&= \frac{I}{\lmin - I} + O\parenv*{\frac{1}{f(n)} + 
	\frac{f(n)}{\log(n)}} \\
	&= \frac{1-\lambda}{\lambda} + O\parenv*{\frac{1}{f(n)} + 
	\frac{f(n)}{\log(n)}}, 
\end{align*}
where the second to last equality is justified by $\frac{1}{1-x} = 1 + 
x + \frac{x^2}{1-x}$ for $x\neq 1$, and since from 
\cref{eq:I-def,eq:lmin-lover} $\frac{I}{\lmin - I} = O(1)$. Finally, 
\begin{align*}
	\ell &= \frac{\lover-2f(n)-6}{1 + \frac{1-\lambda}{\lambda} + 
	O\parenv*{\frac{1}{f(n)} + \frac{f(n)}{\log(n)}}} + O(1) \\
	&= \frac{\lambda \lover - O(f(n))}{1 + 
	O\parenv*{\frac{1}{f(n)}+\frac{f(n)}{\log(n)}}} + O(1) \\
	&= \parenv*{\lambda \lover - O(f(n))} \parenv*{1 - 
	O\parenv*{\frac{1}{f(n)}+\frac{f(n)}{\log(n)}}} \\
	&= \lambda \lover - O\parenv*{\frac{\log(n)}{f(n)} + f(n)}, 
\end{align*}
where again the second to last equality is based on $\frac{1}{1-x} = 
1 + x + \frac{x^2}{1-x}$. \hfill\IEEEQEDhere
\end{IEEEproof}

Based on this property, we show that \cref{cnst:overlap} 
asymptotically meets the bound of~\cref{lem:over-lin-rate}. 
\begin{theorem}\label{thm:single-red}
Letting $f(n)\deq \ceilenv[\Big]{\sqrt{\log(n)}}$, the use 
of~\cref{cor:rf} in~\cref{cnst:overlap} is justified, and we have 
\begin{align*}
	R\parenv*{\code{cnst:overlap}(n)} 
	\geq \frac{1-1/a}{1-\gamma} - 
	\frac{(\log(n))^\epsilon}{a \sqrt{\log(n)}} 
	- O\parenv*{\frac{1}{\sqrt{\log(n)}}}.
\end{align*}
\end{theorem}
\begin{IEEEproof}
We start by noting from \cref{eq:I-def,eq:lmin-lover,eq:r-def} 
\begin{align*}
	\frac{I}{\lmin} &= \frac{\frac{1-\gamma a}{1-\gamma} + 
	(\log(n))^{\epsilon-0.5} + O\parenv*{\frac{1}{\log(n)}}}{a + 
	O\parenv*{\frac{1}{\log(n)}}} \\
	&= \frac{\frac{1-\gamma a}{1-\gamma} + 
	(\log(n))^{\epsilon-0.5}}{a} \parenv*{1 + 
	O\parenv*{\frac{1}{\log(n)}}} \\
	&= \frac{1/a-\gamma}{1-\gamma} + \frac{(\log(n))^\epsilon}{a 
	\sqrt{\log(n)}} + O\parenv*{\frac{1}{\log(n)}}, 
\end{align*}
and 
\begin{align*}
	\frac{r}{\lmin} &= \frac{I \parenv*{1 + O\parenv*{\frac{1}{f(n)} + 
	\frac{f(n)}{\log(n)}}}}{\lmin} \\
	&= \frac{I}{\lmin} + O\parenv*{\frac{1}{f(n)} + 
	\frac{f(n)}{\log(n)}} \\
	&= \frac{1/a-\gamma}{1-\gamma} + \frac{(\log(n))^\epsilon}{a 
	\sqrt{\log(n)}} + O\parenv*{\frac{1}{\sqrt{\log(n)}}}.
\end{align*}

Now, from \cref{eq:N-def} 
\begin{align*}
	N_i &\geq \floorenv*{n/q^I} - \ceilenv*{n/(q^I \lmin)} r \\
	&\geq q^{-I} n \parenv*{1 - r/\lmin} - (r+1) \\
	&= q^{-I} n \parenv*{1 - \tfrac{r}{\lmin} - \tfrac{q^I (r+1)}{n}} 
	= \Omega(q^{-I} n),
\end{align*}
hence 
\begin{align*}
	\log(N_i) &\geq \log(n)-I + O(1) \\
	&= \frac{(a-1) \gamma}{1-\gamma} \log(n) - 
	(\log(n))^{0.5+\epsilon} + O(1).
\end{align*}
In particular for sufficiently large~$n$ we have 
\begin{align}\label{eq:single-red-2}
	f(n) \geq \ceilenv*{\log\log\parenv*{N_i}} + 5.
\end{align}

Next, by~\cref{lem:overlap-prop}, 
\begin{align*}
	\ell &= \parenv*{1-\frac{I}{\lmin}} \lover - 
	O\parenv*{f(n) + \frac{\log(n)}{f(n)}} \\
	&= \parenv*{\frac{1-1/a}{1-\gamma} - 
	\frac{(\log(n))^\epsilon}{a \sqrt{\log(n)}}} \gamma a \log(n) - 
	O\parenv*{\sqrt{\log(n)}} \\
	&= \frac{(a-1) \gamma}{1-\gamma} \log(n) - 
	\gamma (\log(n))^{0.5+\epsilon} - O\parenv*{\sqrt{\log(n)}}.
\end{align*}
Hence 
\begin{align*}
	\ell - \ceilenv*{\log\parenv*{N_{n,\ell}(m)}} &= 
	(1-\gamma) (\log(n))^{0.5+\epsilon} - O\parenv*{\sqrt{\log(n)}} \\
	&> 3 f(n) 
	\label{eq:single-red-1} \numberthis
\end{align*}
for sufficiently large~$n$. Together, 
\cref{eq:single-red-1,eq:single-red-2} satisfy the conditions 
of~\cref{cor:rf}, allowing us to efficiently encode $\bfy_i 
= E^{\rf}_{N_i,\ell}(\bfx_i)\in \rf_\ell\parenv*{N_i}$ (and vice 
versa, decode $\bfx_i$) while attaining from \cref{eq:m} 
$\frac{m(N_i)}{N_i}\geq 1 - \frac{f(n)+2}{N_i} - 
\frac{q^3}{q-2}q^{-f(n)} = 1 - O(q^{-f(n)})$, where the coefficient of 
the asymptotic notation does not depend on~$i$. Hence,
\begin{align*}
	\sum_{i\in [q^I]} m(N_i) 
	&\geq \parenv*{1 - O(q^{-f(n)})} \sum_{i\in [q^I]} N_i \\
	&= \parenv*{1 - O(q^{-f(n)})} 
	\parenv*{n - q^I \ceilenv*{n/(q^I \lmin)} r} \\
	&\geq n \parenv*{1 - O(q^{-f(n)})} 
	\parenv*{1 - \frac{r}{\lmin} - \frac{q^I r}{n}}, 
\end{align*}
where the equality on the second line follows from~\cref{eq:N-def}, 
which concludes the proof.
\end{IEEEproof}
From the proof of~\cref{thm:single-red} we note that $\epsilon$ 
in~\cref{cnst:overlap} must satisfy $\epsilon\geq 
\max\bracenv*{\frac{\log(f(n))}{\log\log(n)}, 
1-\frac{\log(f(n))}{\log\log(n)}}-0.5$; it follows that the choice 
$f(n) = \ceilenv[\big]{\sqrt{\log(n)}}$ is optimal, in the sense that 
$\frac{\log(f(n))}{\log\log(n)} = \frac{1}{2} + o(1)$.

\section{Multi-strand reconstruction from substring-compositions}
\label{sec:multi}

In this section, we study an extension of the reconstruction from 
substring-compositions problem, i.e., $(\ell, \ell-1)$-trace codes, 
to multisets of strings, i.e, to codes over $\cX_{n, k}$ for $k>1$. 
For a string~$\bfx\in \Sigma^n$ we denote for brevity an $(\ell, 
\ell-1)$-trace of~$\bfx$, $\cT_\ell^{\ell-1}(\bfx)$ and 
$\cL_\ell^{\ell-1}(\bfx)$ by an $\ell$-trace, $\cT_\ell(\bfx)$ and 
$\cL_\ell(\bfx)$, respectively. We say $\cL_\ell(\bfx)$ in particular 
is the \emph{$\ell$-profile} of~$\bfx$, the multiset of all of its 
$\ell$-substrings.
 
We shall assume throughout in asymptotic analysis that as $n$ grows, 
$\limsup\frac{\log(k)}{n} < 1$, which is most relevant in 
applications (see, e.g., \cite{Eis20} for an overview of typical 
string-lengths in applications); the complement case is of independent 
theoretical interest, and is left for future work. Hence, we have the 
following lemma.
\begin{lemma}\label{lem:channel}
$\log\abs*{\cX_{n,k}} = k (n - \log(k/e)) + o(k) = \Theta(n k)$. 
\end{lemma}
\begin{IEEEproof}
From Stirling's approximation we have $(k/e)^k \leq k!\leq e \sqrt{k} 
(k/e)^k$, implying 
\begin{IEEEeqnarray*}{+rCl+x*}
	\frac{1}{e\sqrt{k}} \parenv*{\frac{q^n}{k/e}}^k 
	\leq \frac{q^{n k}}{k!}\leq \abs*{\cX_{n,k}} 
	&\leq& \frac{(q^n+k)^k}{k!} \\
	&\leq& \parenv*{\frac{q^n (1+k/q^n)}{k/e}}^k.
	\\[-\normalbaselineskip] &&&\IEEEQEDhere
\end{IEEEeqnarray*}
\end{IEEEproof}

For a \emph{multi-strand $\ell$-trace code}~$\cC$ we have 
from~\cref{lem:code-size} that $\abs*{\cC}\leq 
\binom{n k + q^{\ell}}{q^{\ell}}$. A corollary 
of~\cref{lem:over-lin-rate} is therefore stated:
\begin{corollary}\label{cor:zero-rate}
Assume $\limsup\frac{\log(k)}{n}<1$. If $\log(n k) - \ell = 
\omega_{n k}(1)$ then for any multi-strand $\ell$-trace 
code~$\cC\subseteq \cX_{n, k}$ it holds that 
\begin{align*}
	R(\cC) = o_{n k}(1).
\end{align*}
\end{corollary}
\begin{IEEEproof}
By the observation in the proof of \cref{lem:over-lin-rate} 
\begin{IEEEeqnarray*}{+rCl+x*}
	R(\cC) &\leq& 
	\frac{1}{\abs*{\cX_{n,k}}} \log\binom{n k + q^{\ell}}{q^{\ell}} \\
	&=& O\parenv*{\frac{q^\ell}{n k} 
	\parenv*{2\log(e) + \log(n k) - \ell}}, 
\end{IEEEeqnarray*}
where the equality follows from \cref{lem:channel}.
\end{IEEEproof}

On the other hand, recall from~\cref{cor:reconst-rf-k} that 
$\rf_\ell(n,k)\subseteq \cX_{n, k}$ is a multi-strand $(\ell+1)$-trace 
code. 
Next, we show in contrast to~\cref{cor:zero-rate} that if $\ell - 
\log(nk) = \omega_{n k}(1)$, then $R(\rf_\ell(n, k)) = 1 - o_{nk}(1)$. 
We shall do so by presenting two explicit constructions of 
multi-strand $\ell$-repeat-free codes with efficient encoders and 
decoders. 
For convenience, we assume all quantities to have integer values; a 
straightforward adjustment of the described methods applies for all 
values.

\subsection{Index-based construction}

\begin{construction}\label{cnst:naive}
Denote $n'\deq (n-\log(k)) k$, and take $m$ such that $E\colon 
\Sigma^m\to \rf_{\ell'}(n')$ is any repeat-free encoder, for a given 
$\ell'$. Let $\bfx\in \Sigma^m$ be an arbitrary information string, 
and encode it into $\bfy\deq E(\bfx)$. Take $\bfy_0,\ldots, \bfy_{k-1} 
\in \Sigma^{n-\log(k)}$ such that $\bfy = \bfy_0\circ \bfy_1\circ 
\cdots\circ \bfy_{k-1}$. Let $\bfc_i\in \Sigma^{\log(k)}$ be a $q$-ary 
expansion of $i\in [k]$. Denote $\widetilde{\bfy}_i\deq \bfc_i\circ 
\bfy_i$; then, 
\begin{align*}
	\enc[cnst:naive](\bfx) \deq 
	\bracenv*{\mathset*{\widetilde{\bfy}_i}{i\in [k]}}\in \cX_{n,k}.
\end{align*}

\vspace{-1.5\normalbaselineskip}
\end{construction}

We denote $\code{cnst:naive}(n,k)\deq \enc[cnst:naive](\Sigma^m)$. 
The decoding success of \cref{cnst:naive} follows from the next lemma. 
\begin{lemma}
$\code{cnst:naive}(n,k)\subseteq \rf_\ell(n,k)$, where $\ell = \ell'+
\log(k)$.
\end{lemma}
\begin{IEEEproof}
For $\bfx\in \Sigma^m$, note that $\bfy\deq \enc[cnst:naive](\bfx) = 
\bfy_0\circ \bfy_1\circ \cdots\circ \bfy_{k-1}\in \rf_{\ell'}(n')$ and 
thus $\norm{\cL_{\ell'}(\bfy)} = n'-\ell'+1$. It follows that 
$\norm{\cL_{\ell'}\parenv*{\bracenv*{\mathset*{\bfy_i}{i\in [k]}}}} = 
(n'-\ell'+1) - (k-1) (\ell'-1) = k (n - \ell + 1)$.

Now, let $\bfu,\bfv$ be $\ell$-substrings of $\widetilde{\bfy}_i, 
\widetilde{\bfy}_j$ respectively; note that the $\ell'$-suffixes of 
$\bfu,\bfv$ are $\ell'$-substrings of $\bfy_i, \bfy_j$ respectively, 
and hence if $\bfu=\bfv$ then $i=j$ and their locations in $\bfy_i$ 
agree. It follows that the locations of $\bfu,\bfv$ in 
$\widetilde{\bfy}_i$ agree as well, and the claim follows.
\end{IEEEproof}

Recall, then, that given $\cL_{\ell+1}(\enc[cnst:naive](\bfx))$, an 
efficient algorithm produces the set of strings 
$\mathset{\widetilde{\bfy}_i}{i\in [k]}$. Then, by ordering and 
subsequent removal of the length-$\log(k)$ indices from these strings, 
one obtains the string $\bfy = E(\bfx)$, and consequently, $\bfx$. 
Note that the role of the indices in this construction is crucial to 
deduce~$\bfy$ from its $\ell$-profile; without indices the order of 
these $k$~substrings could not have been derived, hence one would only 
obtain $\bfy$ up to a permutation of its non-overlapping 
$(n'/k)$-substrings. The next theorem analyzes the parameters of codes 
that can be constructed using \cref{cnst:naive} based upon 
\cref{lem:elishco,cor:rf}.

\begin{theorem}\label{thm:rate-naive}
Given $\ell(n,k)$, denote $f(n,k)\deq \ell(n,k)-\log(n k)-\log(k)$. 
Further, let $\ell'\deq \ell(n,k)-\log(k)$. Here, we assume 
\cref{cnst:naive} is operated with $n,k,\ell'$. Observe 
\begin{align*}
	\ell' - \log(n') &= f(n,k) - \log\parenv*{1-\frac{\log(k)}{n}} 
	\geq f(n,k).
\end{align*}
\begin{enumerate}
\item 
If $f(n,k) \geq 3\log\log(n k) + 12$ then utilizing \cref{cor:rf} 
in~\cref{cnst:naive} we obtain 
\begin{align*}
	R(\code{cnst:naive}(n,k)) 
	&\geq 1 - \frac{q^{4 - \floorenv*{f(n,k)/3}}}{q-2} - 
	\frac{\log(e)}{n-\log(k)} - o\parenv*{\tfrac{1}{n}}.
\end{align*}
(For $q=2$ that is $R(\code{cnst:naive}(n,k)) \geq 1 - 
2^{5 - \floorenv*{f(n,k)/3}} - \frac{\log(e)}{n-\log(k)} - 
o\parenv*{\tfrac{1}{n}}$.)

\item \label{it:rate-naive-2}
If $f(n,k)\geq \log(n k) + 2 + 2\log\parenv*{1-\frac{\log(k)}{n}}$ 
then utilizing \cref{lem:elishco} in~\cref{cnst:naive} we have 
\begin{align*}
	R(\code{cnst:naive}(n,k)) 
	&\geq 1 - \frac{\log(e)}{n-\log(k)} - o\parenv*{\tfrac{1}{n}}.
\end{align*}
\end{enumerate}
\end{theorem}
\begin{IEEEproof}
\begin{enumerate}
\item 
Note that by the assumption, \cref{cor:rf} may be applied for some 
choice of~$t$. Since \cref{cnst:naive} does not require $\bfy$ to be 
run-length constrained, we let $t\deq \floorenv*{(\ell'-\log(n'))/3}$ 
and observe 
\begin{align*}
	m &\geq n' - t - 1 - 
	\ceilenv*{\tfrac{q^4}{q-2} n' \big/ q^t} \\
	&= n' - \ceilenv*{\tfrac{q^4}{q-2} n' \big/
	q^{\floorenv*{\parenv*{\ell' - \log(n')}/3}}} \\
	&\geq \parenv*{1 - \tfrac{q^4}{q-2} 
	q^{- \floorenv*{f(n,k)/3}}} k (n-\log(k)) - 1.
\end{align*}
(For $q=2$, that is $m\geq \parenv*{1 - 2^{5 - \floorenv*{f(n,k)/3}}} 
k (n-\log(k)) - 3$.)

It then follows from~\cref{lem:channel} that 
\begin{align*}
	R(\code{cnst:naive}(n,k)) &= \frac{m}{\log\abs*{\cX_{n,k}}} \\
	&\geq \frac{\parenv*{1 - \tfrac{q^{4-\floorenv*{f(n,k)/3}}}{q-2}} 
	k (n-\log(k)) - 1}{k (n - \log(k/e)) + o(k)} \\
	&= \frac{1 - \frac{q^{4 - \floorenv*{f(n,k)/3}}}{q-2} - 
	\frac{1}{(n-\log(k)) k}}{1 + \frac{\log(e)+o(1)}{n-\log(k)}} \\
	&= 1 - \frac{q^{4 - \floorenv*{f(n,k)/3}}}{q-2} - 
	\frac{\log(e)}{n-\log(k)} - o\parenv*{\tfrac{1}{n}}, 
\end{align*}
where again the last equality follows from $\frac{1}{1-x} = 1 + x + 
\frac{x^2}{1-x}$, and from the observation $n-\log(k) = \Theta(n)$. 
(Similarly for $q=2$.)

\item 
Equivalently, $\ell'\geq 2\log(n')+2$, hence by~\cref{lem:elishco} we 
have $m = n'-1$. Following the same steps as in the last part, 
\begin{IEEEeqnarray*}{+rCl+x*}
	R(\code{cnst:naive}(n,k)) &=& \frac{m}{\log\abs*{\cX_{n,k}}} \\
	&\geq& 1 - \frac{\log(e)}{n-\log(k)} - o\parenv*{\tfrac{1}{n}}.
	\\[-\normalbaselineskip] &&&\IEEEQEDhere
\end{IEEEeqnarray*}
\end{enumerate}
\end{IEEEproof}

\subsection{Overlap-based construction}

While in \cref{cnst:naive} we added indices in order to overcome the 
lack of ordering when the string $\bfy =E(\bfx)$ is partitioned 
into~$k$ substrings, in \cref{cnst:sagi} we tackle this constraint 
differently. To wit, we again partition $\bfy$, but include 
overlapping segments between consecutive substrings. The overlapping 
segments will guarantee in decoding that, given the set of 
$k$~substrings, there will be a unique way to concatenate them into 
one long string. As opposed to \cref{cnst:naive}, this approach 
eliminates the need to decrease the length used for repeat-free 
encoders with respect to that of the read substrings, i.e., $\ell$.
\begin{construction}\label{cnst:sagi}
For a given $\ell$, denote $n'\deq n k - (k-1) \ell = (n-\ell) k + 
\ell$, and take $m$ such that $E\colon \Sigma^m\to \rf_\ell(n')$ is 
any repeat-free encoder. Let $\bfx\in \Sigma^m$ be an arbitrary 
information string, and encode it into $\bfy\deq E(\bfx)$. Define $k$ 
length-$n$ strings $\bfy_0, \ldots, \bfy_{k-1}\in \Sigma^n$ by 
segmenting $\bfy$ with an overlap of $\ell$ symbols between 
consecutive segments; more precisely, let $\bfy_i\deq 
\parenv*{y_{i,1},\ldots,y_{i,n}}$ for $i\in [k]$, where 
\begin{align*}
    y_{i,j} &\deq y_{i(n-\ell)+j}; \quad j\in [n].
\end{align*}
Then, 
\begin{align*}
\enc[cnst:sagi](\bfx) \deq 
\bracenv*{\mathset*{\bfy_i}{i\in [k]}}\in \cX_{n,k}.
\end{align*}

\vspace{-1.5\normalbaselineskip}
\end{construction}

We denote $\code{cnst:sagi}(n,k)\deq \enc[cnst:sagi](\Sigma^m)$. 
The decoding success of \cref{cnst:sagi} follows from the following 
simple observation.
\begin{lemma}\label{lem:overlap}
For all $\bfx\in\Sigma^m$ it holds that 
$\cL_{\ell+1}(\bfy) = \cL_{\ell+1}\parenv*{\enc[cnst:sagi](\bfx)}$.
\end{lemma}
\begin{IEEEproof}
Since $\bfy_i$ is a substring of $\bfy$ for all $i$, it follows that 
$\cL_{\ell+1}\parenv*{\enc[cnst:sagi](\bfx)}\subseteq 
\cL_{\ell+1}(\bfy)$. 
For the other direction, note that $\bfy_i,\bfy_{i+1}$ are overlapping 
substrings of $\bfy$ for all $1\leq i<k$, with a common substring of 
length $\ell$; thus all $(\ell+1)$-substrings of $\bfy$ are also 
substrings of some $\bfy_i$.
\end{IEEEproof}

\cref{lem:overlap} immediately implies the next corollary.
\begin{corollary}
$\code{cnst:sagi}(n,k)\subseteq \rf_\ell(n,k)$.
\end{corollary}
\begin{IEEEproof}
By~\cref{lem:overlap} and since $\bfy\in \rf_\ell(n')$.
\end{IEEEproof}

We are now ready to analyze the code parameters that \cref{cnst:sagi} 
can achieve, again based on \cref{lem:elishco,cor:rf}.

\begin{theorem}\label{thm:rate-sagi}
Given $\ell(n,k)$, denote $f(n,k)\deq \ell(n,k)-\log(n k)$. 
\begin{enumerate}
\item 
If $f(n,k) \geq 3\log\log(n k) + 12$ then utilizing \cref{cor:rf} 
in~\cref{cnst:sagi} we obtain 
\begin{IEEEeqnarray*}{+rCl+x*}
	R(\code{cnst:sagi}(n,k)) 
	&\geq& 1 - \frac{q^{4 - \floorenv*{f(n,k)/3}}}{q-2} \>- \\
	&& -\> (1+o(1)) \frac{\log(n) + f(n,k)}{n - \log(k)}.
\end{IEEEeqnarray*}
(for $q=2$, that is $R(\code{cnst:sagi}(n,k)) \geq 1 - 
2^{5 - \floorenv*{f(n,k)/3}} - 
(1+o(1)) \frac{\log(n) + f(n,k)}{n - \log(k)}$).

\item 
If $f(n,k)\geq \log(n k) + 2 + 2\log\parenv*{1-\parenv*{1-\frac{1}{k}} 
\frac{\ell(n,k)}{n}}$ then utilizing 
\cref{lem:elishco} in~\cref{cnst:sagi} we have 
\begin{align*}
	R(\code{cnst:sagi}(n,k)) 
	&\geq 1 - (1+o(1)) \frac{\log(n) + f(n,k)}{n - \log(k)}.
\end{align*}
\end{enumerate}
\end{theorem}
\begin{IEEEproof}
Recalling from~\cref{cnst:sagi} that $n' = n k - (k-1) \ell(n,k)$, we 
begin by observing 
\begin{align*}
	\frac{n'}{n k} &= 1 - \parenv*{1-\frac{1}{k}} \frac{\ell(n,k)}{n} \\
	&= 1 - \frac{\log(k)}{n} - (1+o(1)) \frac{\log(n) + f(n,k)}{n}, 
\end{align*}
hence $\ell(n,k) - \log(n') = \ell(n,k) - \log(n k) + O(1) = f(n,k) + 
O(1)$. Also, by multiplying the above equality with $\frac{n}{n-
\log(k)}$ we have
have 
\begin{align*}
	\frac{n'}{(n-\log(k)) k} &= 1 - 
	(1+o(1)) \frac{\log(n) + f(n,k)}{n - \log(k)}.
\end{align*}

Next, 
\begin{enumerate}
\item 
As in the proof of~\cref{thm:rate-naive}, 
\begin{align*}
	m &\geq \parenv*{1 - \tfrac{q^4}{q-2} 
	q^{- \floorenv*{f(n,k)/3}}} n' - 1, 
\end{align*}
and following the same steps  
\begin{IEEEeqnarray*}{+rCl+x*}
	R(\code{cnst:sagi}(n,k)) &=& \frac{m}{\log\abs*{\cX_{n,k}}} \\
	&\geq& 1 - \frac{q^{4 - \floorenv*{f(n,k)/3}}}{q-2} \>- \\
	&& -\> (1+o(1)) \frac{\log(n) + f(n,k)}{n - \log(k)}.
\end{IEEEeqnarray*}

\item 
Again, we have $\ell(n,k)\geq 2\log(n')+2$, hence $m = n'-1$. It 
follows that 
\begin{IEEEeqnarray*}{+rCl+x*}
	R(\code{cnst:sagi}(n,k)) 
	&=& 1 - (1+o(1)) \frac{\log(n) + f(n,k)}{n - \log(k)}.
	\\[-\normalbaselineskip] &&&\IEEEQEDhere
\end{IEEEeqnarray*}
\end{enumerate}
\end{IEEEproof}

We note that inherent to \cref{cnst:sagi} is that the last step might 
introduce more redundancy than is required for repeat-free encoding. 
Indeed, for $f(n,k)\geq 3 \log(n)$ the latter term 
in~\cref{thm:rate-sagi} becomes significant, and the construction's 
rate is then correspondingly decreasing in $f(n,k)$; this is an oddity 
since $\rf_{\ell_1}(n,k)\subseteq \rf_{\ell_2}(n,k)$ for all $\ell_1 
\leq \ell_2$.

\subsection{Constructions' rates}

In this section we study the performance of the two proposed 
constructions. 
We first seek to give a converse to \cref{cor:zero-rate} and establish 
the result on the minimum value of $\ell$ which guarantees that the 
asymptotic rate of multi-strand $\ell$-reconstruction codes (in fact, 
$\rf_{\ell-1}(n,k)$) is $1$. This result is established in the next 
corollary using~\cref{cnst:sagi}.

\begin{corollary}\label{cor:rate-one}
For $n,k$ satisfying $\limsup\frac{\log(k)}{n}<1$ and for $\ell \geq 
\log(n k) + 3\log\log(n k) + 12$, it holds that  $R(\rf_{\ell}(n,k)) 
= 1-o_{nk}(1).$
\end{corollary}

\begin{figure}[t]{}%
\centering
\includegraphics[width=0.95\columnwidth]{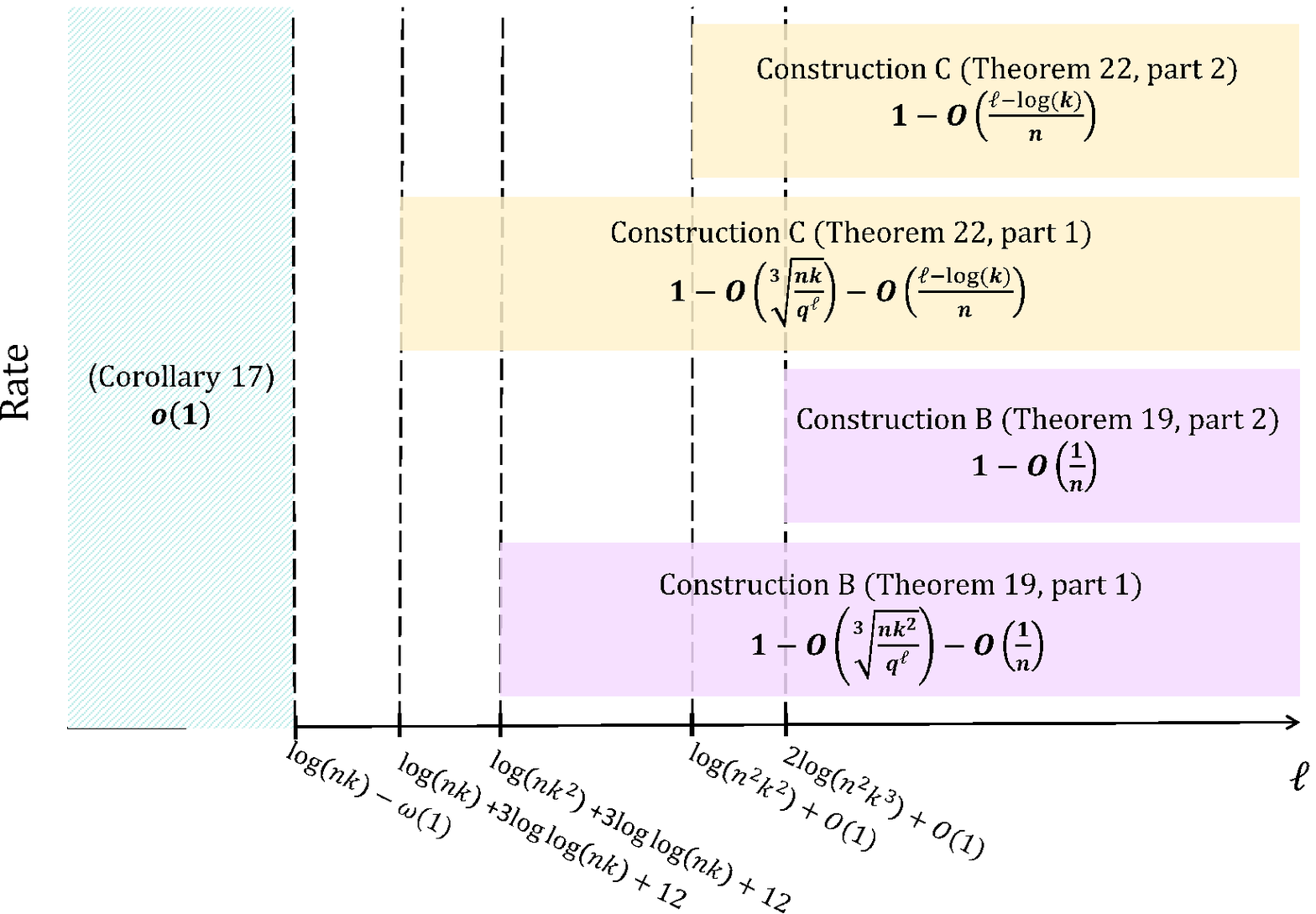}%
\caption[Rates]{Trade-off of window-length to constructions' rates.
\label{fig:compare}}
\end{figure}

Note that if one aims to achieve rate $1-o(1)$ using 
\cref{cnst:naive}, then the minimum value of $\ell(n,k)$ should be 
$\log(n k^2) + 3\log\log(n k) + 12$, i.e., there exists a gap of 
$\log(k)$ with respect to the result in~\cref{cor:rate-one}. However, 
for comparable values of $\ell(n,k)$, \cref{cnst:naive} may offer 
better code rate; a comparison of the rates of both constructions, 
based on~\cref{thm:rate-naive,thm:rate-sagi}, for applicable values 
of~$\ell(n,k)$ is illustrated in~\cref{fig:compare}, in context of the 
lower bound of~\cref{cor:zero-rate}. The following 
observation follows from these results.

\begin{lemma}
$R(\code{cnst:naive}(n)) > R(\code{cnst:sagi}(n))$ for sufficiently 
large~$n$ if 
\begin{enumerate}
\item \label{part:1} 
$\ell(n,k)\geq \log(n^2 k^3)+2+2\log\parenv*{1-\frac{\log(k)}{n}}$; or

\item \label{part:2} 
if $k=\Omega(n^2)$, for $\ell - \log(n^4 k^2) + 3\log\log(n^4 k) = 
\omega(1)$; or 

\item \label{part:3} 
if $\log(k) = \omega(\sqrt{n})$, for $\ell\geq \log(n k^2) + 
3\log\log(n k) + 12$.
\end{enumerate}
\end{lemma}
\begin{IEEEproof}
Clearly the claim holds if $\ell(n,k)\geq \log(n^2 k^3) + 2 + 
2\log\parenv*{1-\frac{\log(k)}{n}}$ by~\cref{it:rate-naive-2} 
of~\cref{thm:rate-naive}, satisfying \cref{part:1}.

For lower values of~$\ell = \ell(n,k)$, suffice that 
$\frac{\ell-\log(k)}{n} = \omega\parenv*{\sqrt[3]{\frac{n 
k^2}{q^\ell}}}$. Reorganizing $q^{(\ell-\log(k))/3} (\ell-\log(k))/3 = 
\omega\parenv*{\sqrt[3]{n^4 k}}$, we equivalently have 
$\frac{\ln(q)}{3} (\ell-\log(k)) = W_0\parenv*{\sqrt[3]{n^4 k}} + 
\omega(1)$, where $W_0(x) = \ln(x)-\ln\ln(x)+o(1)$ is the principal 
brunch of the Lambert W function. Hence, a sufficient condition is 
that 
\begin{align}\label{eq:ell-B-best}
	\ell - \log(n^4 k^2) + 3\log\log(n^4 k) = \omega(1).
\end{align}

For \cref{part:2}, observe that 
\begin{IEEEeqnarray*}{+rCl+x*}
	\IEEEeqnarraymulticol{3}{l}{%
	\parenv*{\log(n^2 k^3) + 2 + 2\log\parenv*{1-\frac{\log(k)}{n}}}} \\
	\IEEEeqnarraymulticol{3}{r}{%
	-\> \log(n^4 k^2) + 3\log\log(n^4 k)} \\
	&=& \log(k/n^2) + 3\log\log(n^4 k) + O(1), 
\end{IEEEeqnarray*}
hence $k=\Omega(n^2)$ ensures that there exist values of~$\ell$ 
satisfying both \cref{eq:ell-B-best} and $\ell < \log(n^2 k^3) + 2 
+ 2\log\parenv*{1-\frac{\log(k)}{n}}$, i.e., not already covered 
by~\cref{part:1}.

Finally, \cref{part:3} is justified by 
\begin{IEEEeqnarray*}{+rCl+x*}
	\IEEEeqnarraymulticol{3}{l}{%
	\parenv*{\log(n k^2) + 3\log\log(n k) + 12}} \\
	\IEEEeqnarraymulticol{3}{r}{%
	-\> \log(n^4 k^2) + 3\log\log(n^4 k)} \\
	&=& 3\log\parenv*{\frac{\log(n k) \log(n^4 k)}{n}} + 12, 
\end{IEEEeqnarray*}
and the observation that 
$\log(n k) \log(n^4 k) = \omega(n)$ if and only if $\log(k) = 
\omega(\sqrt{n})$.
\end{IEEEproof}

\section{Conclusion}\label{sec:conc}

In this work, we generalized both the reconstruction from 
substring-composition problem, and the torn-paper problem, by studying 
an intermediate setting of partial overlap between read substrings. 
Our analysis is done in worst-case (i.e., adversarial) regime, as 
opposed to the probabilistic treatment of this problem 
in~\cite{RavVahSho22}. For the case of a single string ($k=1$), we 
proved an upper bound on achievable code rates (implying in particular 
a lower bound on the length of read substrings, required for 
asymptotically non-vanishing codes' rates), and developed an efficient 
construction asymptotically achieving optimal rate. 
Pleasingly, at the two extreme points, \cref{cnst:overlap} essentially 
degenerates to known constructions for either the torn-paper 
channel~\cite{BarMarYaaYeh22} (for $\gamma = 0$) or for reconstruction 
from substring-composition~\cite{EliGabMedYaa21} (for $\gamma\to 
1/a$). 
Finally, we demonstrate that like in the torn-paper extreme, one may 
also extend solutions to the reconstruction from substring-composition 
problem to multiset-codes. It is left for future work to extend the 
intermediate setting in this fashion.

Before concluding, we suggest that one might consider a slightly 
different channel definition to that of \cref{sec:multi}, where the 
$k$~strands are required to be distinct from one another, i.e., when 
information is stored in the space 
\begin{align*}
	\cX^*_{n,k} \deq \mathset*{S\subseteq \Sigma^n}{\abs*{S}=k}.
\end{align*}
A priori, it seems feasible that the added restriction might allow 
for lower redundancy (when measured in $\cX^*_{n,k}$). However, we 
note that $\abs*{\cX^*_{n,k}} = \binom{q^n}{k}$, thus a similar 
development to \cref{lem:channel} yields 
\begin{align*}
	\frac{(q^n-k)^k}{k!} \leq \abs*{\cX^*_{n,k}} 
	\leq \frac{q^{n k}}{k!}.
\end{align*}
It follows that $\log\abs*{\cX^*_{n,k}} = k (n - \log(k/e)) + o(k)$ as 
well. A careful examination reveals that \cref{cnst:naive,cnst:sagi} 
actually encode into $\cX^*_{n,k}\cap \rf_\ell(n,k)$, and hence the 
results of this work also hold for that setup of the problem.

\section*{Acknowledgments}

The authors gratefully acknowledge the two anonymous reviewers, and 
associate editor, whose insight and suggestions helped shape 
this paper and greatly improve its presentation.




\begin{IEEEbiographynophoto}{Yonatan Yehezkeally}
(S'12--M'20)
is the Carl Friedrich von Siemens postdoctoral research fellow of the 
Alexander von Humboldt Foundation, in the Associate Professorship of 
Coding and Cryptography (Prof. Wachter-Zeh), School of Computation, 
Information and Technology, Technical University of Munich. 
His research interests include coding for novel storage media, 
with a focus on DNA-based storage and nascent sequencing technologies, 
as well as combinatorial structures and finite group theory.

Yonatan received the B.Sc.~(\emph{cum laude}) degree in Mathematics, 
and the M.Sc.~(\emph{summa cum laude}) and Ph.D. degrees in Electrical 
and Computer Engineering, in 2013, 2017 and 2020 respectively, all 
from Ben-Gurion University of the Negev, Beer-Sheva, Israel.
\end{IEEEbiographynophoto}

\begin{IEEEbiographynophoto}{Daniella Bar-Lev}
(S'20)
is a Ph.D. student in the Computer Science Department at the Technion 
--- Israel Institute of Technology. 
She received the B.Sc. degrees in computer science and mathematics, 
and the M.Sc. degree in computer science from the Technion --- Israel 
Institute of Technology, Haifa, Israel, in~2019 and~2021, 
respectively. Her research interests include algorithms, discrete 
mathematics, coding theory, and  DNA storage.
\end{IEEEbiographynophoto}

\begin{IEEEbiographynophoto}{Sagi Marcovich}
(S'20)
is a Ph.D. student in the Computer Science Department at the Technion 
--- Israel Institute of Technology. He received the B.Sc. degree in 
software engineering and his M.Sc. degree in computer science from the 
Technion --- Israel Institute of Technology, Haifa, Israel in~2016 
and~2021, respectively. 
His research interests include algorithms, information theory, and 
coding theory with applications to DNA based storage.
\end{IEEEbiographynophoto}

\begin{IEEEbiographynophoto}{Eitan Yaakobi}
(S'07--M'12--SM'17) is an Associate Professor at the Computer Science 
Department at the Technion --- Israel Institute of Technology. He also 
holds a courtesy appointment in the Technion's Electrical and Computer 
Engineering (ECE) Department. He received the B.A. degrees in computer 
science and mathematics, and the M.Sc. degree in computer science from 
the Technion --- Israel Institute of Technology, Haifa, Israel, 
in~2005 and~2007, respectively, and the Ph.D. degree in electrical 
engineering from the University of California, San Diego, in~2011. 
Between~2011-2013, he was a postdoctoral researcher in the department 
of Electrical Engineering at the California Institute of Technology 
and at the Center for Memory and Recording Research at the University 
of California, San Diego. His research interests include information 
and coding theory with applications to non-volatile memories, 
associative memories, DNA storage, data storage and retrieval, and 
private information retrieval. He received the Marconi Society Young 
Scholar in~2009 and the Intel Ph.D. Fellowship in~2010-2011. 
Since~2020, he serves as an Associate Editor for Coding snd Decoding 
for the \textsc{IEEE Transactions on Information Theory}. Since~2016, 
he is affiliated with the Center for Memory and Recording Research at 
the University of California, San Diego, and since~2018, he is 
affiliated with the Institute of Advanced Studies, Technical 
University of Munich, where he holds a four-year Hans Fischer 
Fellowship, funded by the German Excellence Initiative and the EU 7th 
Framework Program. He is a recipient os several grants, including the 
ERC Consolidator Grant. 
\end{IEEEbiographynophoto}

\end{document}